\documentclass[12pt,english,draftcls, one column]{IEEEtran}
\usepackage[T1]{fontenc}
\usepackage[latin9]{inputenc}
\usepackage{float}
\usepackage{amsmath}
\usepackage{amsthm}
\usepackage{amssymb}
\usepackage{graphicx}

\makeatletter

\floatstyle{ruled}
\newfloat{algorithm}{tbp}{loa}
\providecommand{\algorithmname}{Algorithm}
\floatname{algorithm}{\protect\algorithmname}

\theoremstyle{plain}

\theoremstyle{plain}

\ifCLASSINFOpdf
\else
\fi
\usepackage{babel}

\makeatother

\usepackage{babel}
\providecommand{\propositionname}{Proposition}
\providecommand{\theoremname}{Theorem}

\begin{document}

\title{Multi-Tenant C-RAN With Spectrum Pooling: Downlink Optimization Under
\\Privacy Constraints}

\author{Seok-Hwan Park, \textit{Member}, \textit{IEEE}, Osvaldo Simeone,
\textit{Fellow, IEEE}, \\ and Shlomo Shamai (Shitz),\textit{ Fellow,
IEEE} \thanks{S.-H. Park was supported by the National Research Foundation of Korea
(NRF) grant funded by the Korea government (Ministry of Science, ICT\&Future
Planning) {[}2015R1C1A1A01051825{]}. The work of O. Simeone was partially
supported by the U.S. NSF through grant 1525629. O. Simeone has also
received funding from the European Research Council (ERC) under the
European Union\textquoteright s Horizon 2020 research and innovation
programme (grant agreement No 725731). The work of S. Shamai has been
supported by the Israel Science Foundation (ISF) and by the ERC Advanced
Grant, No. 694630.

S.-H. Park is with the Division of Electronic Engineering, Chonbuk
National University, Jeonju 54896, Korea (email: seokhwan@jbnu.ac.kr).

O. Simeone is with the Department of Informatics, King\textquoteright s
College London, London, UK (email: osvaldo.simeone@kcl.ac.uk).

S. Shamai (Shitz) is with the Department of Electrical Engineering,
Technion, Haifa, 32000, Israel (email: sshlomo@ee.technion.ac.il).}}
\maketitle
\begin{abstract}
Spectrum pooling allows multiple operators, or tenants, to share the
same frequency bands. This work studies the optimization of spectrum
pooling for the downlink of a multi-tenant Cloud Radio Access Network
(C-RAN) system in the presence of inter-tenant privacy constraints.
The spectrum available for downlink transmission is partitioned into
private and shared subbands, and the participating operators cooperate
to serve the user equipments (UEs) on the shared subband. The network
of each operator consists of a cloud processor (CP) that is connected
to proprietary radio units (RUs) by means of finite-capacity fronthaul
links. In order to enable inter-operator cooperation, the CPs of the
participating operators are also connected by finite-capacity backhaul
links. Inter-operator cooperation may hence result in loss of privacy.
Fronthaul and backhaul links are used to transfer quantized baseband
signals. Standard quantization is considered first. Then, a novel
approach based on the idea of correlating quantization noise signals
across RUs of different operators is proposed to control the trade-off
between distortion at UEs and inter-operator privacy. The problem
of optimizing the bandwidth allocation, precoding, and fronthaul/backhaul
compression strategies is tackled under constraints on backhaul and
fronthaul capacity, as well as on per-RU transmit power and inter-operator
privacy. For both cases, the optimization problems are tackled using
the concave convex procedure (CCCP), and extensive numerical results
are provided.\end{abstract}


\theoremstyle{theorem}
\newtheorem{theorem}{Theorem}
\theoremstyle{proposition}
\newtheorem{proposition}{Proposition}
\theoremstyle{lemma}
\newtheorem{lemma}{Lemma}
\theoremstyle{corollary}
\newtheorem{corollary}{Corollary}
\theoremstyle{definition}
\newtheorem{definition}{Definition}
\theoremstyle{remark}
\newtheorem{remark}{Remark}

\section{Introduction}

Spectrum pooling among multiple network operators, or tenants, is
an emerging technique for meeting the rapidly increasing traffic demands
over the available scarce spectrum resources \cite{Khan-et-al}-\cite{JPark-et-al}.
Spectrum pooling can be implemented by means of orthogonal or non-orthogonal
resource allocation. In orthogonal spectrum pooling, the frequency
channels are exclusively, but dynamically, allocated to the participating
operators \cite{Jorswieck-et-al}. In contrast, with non-orthogonal
spectrum pooling, parts of the spectrum can be shared between operators.
In addition to spectrum pooling, radio access network (RAN) sharing,
whereby RAN infrastructure nodes are shared by the tenants, has also
been considered \cite{Boccardi-et-al}\cite{Aydin-et-al}. RAN sharing
and spectrum pooling are two examples of network slicing, a key technology
for the upcoming 5G wireless systems \cite{Samdanis-et-al}\cite{Foukas-et-al}.

\begin{figure}
\centering\includegraphics[width=13cm,height=8cm]{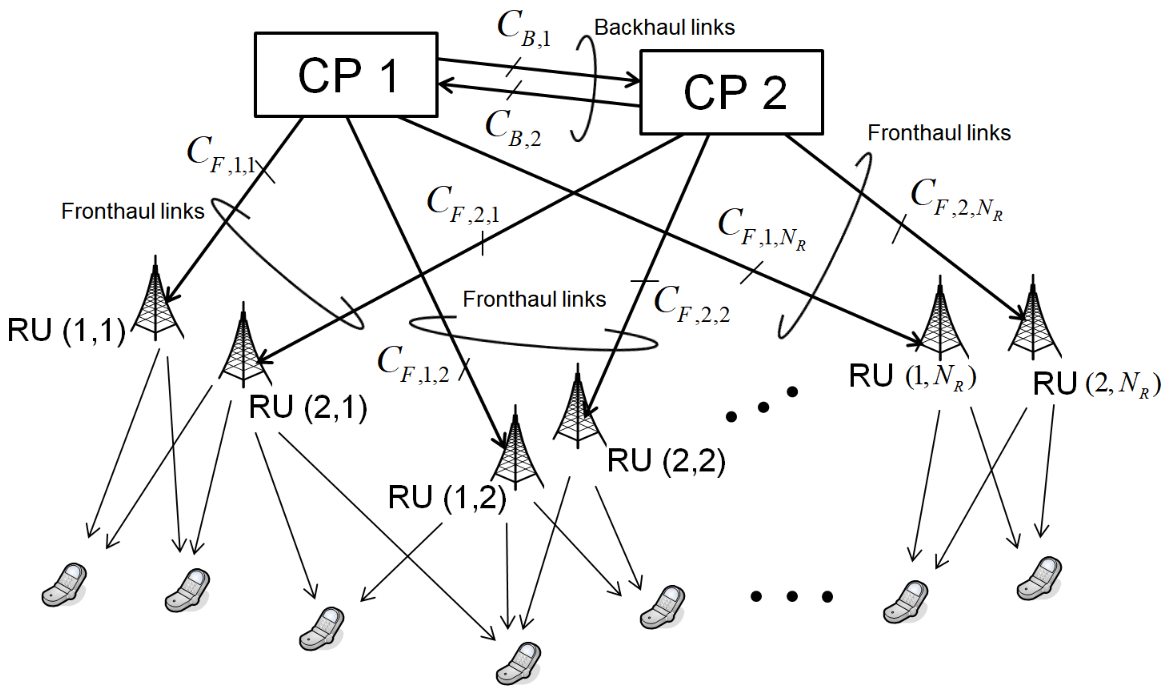}

\caption{\label{fig:System-Model}Illustration of the downlink of a multi-tenant
C-RAN system.}
\end{figure}

In a Cloud RAN (C-RAN) architecture, a Cloud Processor (CP) carries
out centralized baseband signal processing on behalf of a number of
the connected Radio Units (RUs). The CP communicates quantized baseband
signals over fronthaul links, while the RUs only perform radio frequency
functionalities \cite{Simeone-et-al:JCN}\cite{Quek-et-al}. Motivated
by the promised reduction in capital and operational expenditures,
the C-RAN technology is currently being deployed for testing. In this
paper, we focus on the optimization of spectrum pooling across multiple
tenants in a C-RAN architecture, as illustrated in Fig. \ref{fig:System-Model}.

Existing papers on C-RAN downlink optimization, such as \cite{Simeone-et-al:ETT}-\cite{Liu-Yu},
have focused on single-tenant systems. These works, and related references,
study the design of coordinated precoding and fronthaul compression
strategies. Specifically, references \cite{Simeone-et-al:ETT}\cite{Park-et-al:TWC}\cite{Liu-Yu}
consider the use of standard point-to-point fronthaul compression
and quantization strategies, whereas \cite{Park-et-al:TSP13}\cite{Park-et-al:SPM}\cite{Lee-et-al:TSP16}
investigate a more advanced approach based on multivariate compression
and quantization. These perform the joint compression and quantization
of baseband signals across multiple RUs, with the aim of controlling
the impact of quantization distortion at the user equipments (UEs).
Dual approaches for the uplink of C-RAN were studied in, e.g., \cite{Park-et-al:TVT13}\cite{Zhou-et-al:TIT16}.
We refer to \cite{Simeone-et-al:JCN}\cite{Quek-et-al} for a comprehensive
review.

Tackling the optimization of C-RAN systems in the presence of multiple
operators presents novel optimization degrees of freedom and technical
challenges. As a key novel design dimension, the available bandwidth
may be optimally split into private and shared subbands, where the
private subbands are exclusively used by the respective operators
while the shared subband is shared by all the participating operators
(see Fig. \ref{fig:BW-split}). Furthermore, cooperation and coordination
on the shared subband are facilitated by communication between the
CPs, which requires the design of the signals exchanged on the inter-CP
interface as a function of the inter-CP capacity. Finally, the optimization
problem entails a trade-off between the benefits accrued from inter-operator
cooperation and the amount of information exchanged about the respective
users' data.

In this paper, we study the design of the mult-tenant C-RAN system
illustrated in Fig. \ref{fig:System-Model} under the assumption that
fronthaul and inter-CP backhaul links carry quantized baseband signals.
Note that this is the standard mode of operation for C-RAN CP-toRUs
links. We tackle the joint optimization of bandwidth allocation and
of precoding and quantization strategies under constraints on fronthaul
and backhaul capacity and privacy for the inter-CP communications.

To this end, we first consider standard point-to-point quantization
as in most prior work on C-RAN. Then, a novel quantizatoin scheme
based on multivariate compression \cite{Park-et-al:TSP13}\cite{Lee-et-al:TSP16}
is proposed. Through this approach, the CP of an operator is able
to correlate the quantization noise signals across the RUs of \textit{both}
operators, so as to better control the trade-off between the distortion
observed by the UEs and inter-operator privacy. Note that the crucial
element of inter-operator privacy was not present in prior works \cite{Park-et-al:TSP13}\cite{Park-et-al:SPM}\cite{Lee-et-al:TSP16}.
In this regard, we note that the CP and RUs of one operator act as
untrusted relays for the other operators, and hence the proposed technique
can also be applied for the relay channels with untrusted relays studied
in \cite{He-Yener}\cite{Bassily-et-al}. The formulated optimization
problems, albeit non-convex, can be tackled via the concave convex
procedure (CCCP) upon rank relaxation \cite{Park-et-al:TSP13}\cite{Tao-et-al}.

The rest of the paper is organized as follows. The system model is
described in Sec. \ref{sec:System-Model}, and Sec. \ref{sec:Multi-Tenant-C-RAN-With}
presents the operation of the multi-tenant C-RAN system with spectrum
pooling. We discuss the optimization of the multi-tenant C-RAN system
in Sec. \ref{sec:Optimization}. The novel multivariate compression
scheme is introduced in Sec. \ref{sec:Multivariate-Quantization}.
We provide numerical results that validate the advantages of optimized
spectrum pooling and of multivariate compression in Sec. \ref{sec:Numerical-Results},
and the paper is concluded in Sec. \ref{sec:Conclusion}.

Some notations used throughout the paper are summarized as follows.
The mutual information between the random variables $X$ and $Y$
is denoted as $I(X;Y)$, and $h(X|Y)$ denotes the conditional differential
entropy of $X$ given $Y$. We use the notation $\mathcal{CN}(\mbox{\boldmath${\mu}$},\bold{R})$
to denote the circularly symmetric complex Gaussian distribution with
mean $\mbox{\boldmath${\mu}$}$ and covariance matrix $\mathbf{R}$.
The set of all $M\times N$ complex matrices is denoted by $\mathbb{C}^{M\times N}$,
and $\mathtt{E}(\cdot)$ represents the expectation operator. The
operation $(\cdot)^{\dagger}$ denotes Hermitian transpose of a matrix
or vector.

\section{System Model\label{sec:System-Model}}

We consider the downlink of a multi-tenant C-RAN with $N_{O}$ operators.
As shown in Fig. \ref{fig:System-Model}, we focus on the case of
$N_{O}=2$ operators, but the treatment could be generalized for any
$N_{O}$ at the cost of a more cumbersome notation. We assume that
each operator has a single CP, $N_{R}$ RUs and $N_{U}$ UEs. We denote
the $r$th RU and the $k$th UE of the $i$th operator as RU $(i,r)$
and UE $(i,k)$, respectively. We consider a general MIMO set-up in
which RU $(i,r)$ and UE $(i,k)$ have $n_{R,i,r}$ and $n_{U,i,k}$
antennas, respectively, and define the number $n_{R,i}\triangleq\sum_{r\in\mathcal{N}_{R}}n_{R,i,r}$
of total RU antennas of each operator. The sets of RU and UE indices
for either operator are denoted as $\mathcal{N}_{R}\triangleq\{1,2,\ldots,N_{R}\}$
and $\mathcal{N}_{U}\triangleq\{1,2,\ldots,N_{U}\}$, respectively,
while $\mathcal{N}_{O}\triangleq\{1,2\}$ is the set of operator indices.

The CP of each operator $i$, indicated as CP $i$, has a message
$M_{i,k}\in\{1,2,\ldots,2^{nR_{i,k}}\}$ to deliver to UE $(i,k)$,
where $n$ is the coding block length, assumed to be sufficiently
large, and $R_{i,k}$ denotes the rate of the message $M_{i,k}$ in
bits per second (bit/s).

As in related works for C-RAN systems (see, e.g., \cite{Simeone-et-al:ETT}-\cite{Park-et-al:SPM}),
we assume that CP $i$ is connected to RU $(i,r)$ by a \textit{fronthaul
link} of capacity $C_{F,i,r}$ bit/s. In addition, in order to enable
inter-operator cooperation, we assume that, as suggested in \cite{Boccardi-et-al},
the CPs of two operators are connected to each other. Specifically,
CP $i$ can send information to the other CP $\bar{i}$ on a \textit{backhaul
link} of capacity $C_{B,i}$ bit/s, where $\bar{i}$ indicates $\bar{i}=3-i$,
i.e., $\bar{1}=2$ and $\bar{2}=1$. We note that it would be generally
useful to deploy interfaces between the RUs of different operators
\cite{Boccardi-et-al}, but this work focuses on investigating the
advantages of inter-CP connections only.

Inter-operator cooperation via RAN sharing, as enabled by the inter-CP
backhaul links, may cause information leakage from one operator to
the other, which may degrade the confidentiality of the UE messages.
When designing the multi-tenant C-RAN system, we will hence impose
privacy constraints such that the inter-operator information leakage
rate does not exceed a given tolerable threshold value.

We assume flat-fading channel models, and divide the downlink bandwidth
as shown in Fig. \ref{fig:BW-split} into private and shared subbands.
The signal $\mathbf{y}_{i,k}^{(i)}\in\mathbb{C}^{n_{U,i,k}\times1}$
received by UE $(i,k)$ on private subband $i$ can be written as
\begin{equation}
\mathbf{y}_{i,k}^{(i)}=\sum_{r\in\mathcal{N}_{R}}\mathbf{H}_{i,k}^{i,r}\mathbf{x}_{i,r}^{(i)}+\mathbf{z}_{i,k}^{(i)},\label{eq:received-signal-private}
\end{equation}
where $\mathbf{H}_{i,k}^{j,r}\in\mathbb{C}^{n_{U,i,k}\times n_{R,j,r}}$
represents the channel matrix from RU $(j,r)$ to UE $(i,k)$; $\mathbf{x}_{i,r}^{(i)}\in\mathbb{C}^{n_{R,i,r}\times1}$
is the signal transmitted by RU $(i,r)$ on the private subband $i$;
and $\mathbf{z}_{i,k}^{(i)}\sim\mathcal{CN}(\mathbf{0},\mathbf{I})$
denotes the additive noise. Similarly, the signal $\mathbf{y}_{i,k}^{(S)}\in\mathbb{C}^{n_{U,i,k}\times1}$
received by UE $(i,k)$ on the shared subband is given as
\begin{equation}
\mathbf{y}_{i,k}^{(S)}=\sum_{r\in\mathcal{N}_{R}}\mathbf{H}_{i,k}^{i,r}\mathbf{x}_{i,r}^{(S)}+\sum_{r\in\mathcal{N}_{R}}\mathbf{H}_{i,k}^{\bar{i},r}\mathbf{x}_{\bar{i},r}^{(S)}+\mathbf{z}_{i,k}^{(S)},\label{eq:received-signal-shared}
\end{equation}
where $\mathbf{x}_{i,r}^{(S)}\in\mathbb{C}^{n_{R,i,r}\times1}$ is
the signal transmitted by RU $(i,r)$ on the shared subband; and $\mathbf{z}_{i,k}^{(S)}\sim\mathcal{CN}(\mathbf{0},\mathbf{I})$
is the additive noise.

\section{Multi-Tenant C-RAN With Spectrum Pooling\label{sec:Multi-Tenant-C-RAN-With}}

In this section, we describe the operation of the multi-tenant C-RAN
system with spectrum pooling and RAN infrastructure sharing by means
of inter-CP connections.

\subsection{Overview}

\begin{figure}
\centering\includegraphics[width=12.8cm,height=2.7cm]{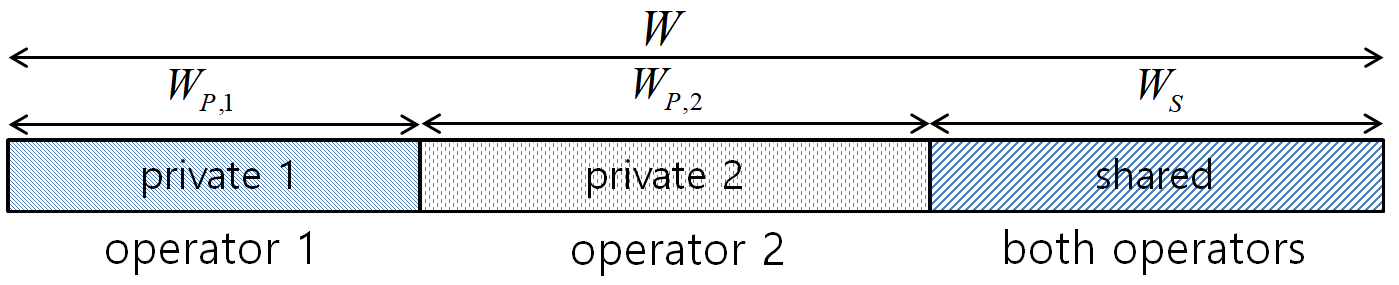}

\caption{\label{fig:BW-split}Illustration of frequency band splitting for
the downlink transmission into private and shared bands.}
\end{figure}

As illustrated in Fig. \ref{fig:BW-split}, we split the frequency
band of bandwidth $W$ {[}Hz{]} into three subbands, where the first
two subbands are exclusively used by the respective operators, while
the last subband is shared by both operators. Accordingly, the bandwidth
$W$ is decomposed as
\begin{equation}
W=W_{P,1}+W_{P,2}+W_{S},\label{eq:bandwidth-splitting}
\end{equation}
where $W_{P,i}$ is the bandwidth of the private subband assigned
to operator $i$, and $W_{S}$ is the bandwidth of the shared subband.

The private subbands are used by each operator to communicate to their
respective UEs with no interference from the other operators' RUs
using standard fronthaul-enabled C-RAN transmission \cite{Simeone-et-al:JCN}.
In contrast, the shared subband is used simultaneously by the two
operators, which can coordinate their transmission through the inter-CP
links. In the following, we detail the operation of CPs, RUs and UEs.

\subsection{Encoding at CPs}

In order to enable transmission over the private and shared subbands,
we split the message $M_{i,k}$ intended for each UE $(i,k)$ into
two submessages $M_{i,k,P}$ and $M_{i,k,S}$ of rates $R_{i,k,P}$
and $R_{i,k,S}$, respectively, with $R_{i,k,P}+R_{i,k,S}=R_{i,k}$.
The submessages $M_{i,k,P}$ and $M_{i,k,S}$ are communicated to
the UE $(i,k)$ on the private and shared subbands, respectively.
Each submessage $M_{i,k,m}$, $m\in\{P,S\}$, is encoded by CP $i$
in a baseband signal $\mathbf{s}_{i,k,m}\in\mathbb{C}^{d_{i,k,m}\times1}$.
We consider standard random coding with Gaussian codebooks, and hence
each symbol $\mathbf{s}_{i,k,m}$ is distributed as $\mathbf{s}_{i,k,m}\sim\mathcal{CN}(\mathbf{0},\mathbf{I})$.

\subsubsection{Linear Precoding for Private Subband}

CP $i$ linearly precodes the signals $\{\mathbf{s}_{i,k,P}\}_{k\in\mathcal{N}_{U}}$
to be transmitted on the private subband as
\begin{equation}
\tilde{\mathbf{x}}_{i}^{(i)}=\left[\tilde{\mathbf{x}}_{i,1}^{(i)\dagger}\,\cdots\,\tilde{\mathbf{x}}_{i,N_{R}}^{(i)\dagger}\right]^{\dagger}=\sum_{k\in\mathcal{N}_{U}}\mathbf{V}_{i,k}^{(i)}\mathbf{s}_{i,k,P},\label{eq:precoding-private}
\end{equation}
where the subvector $\tilde{\mathbf{x}}_{i,r}^{(P)}\in\mathbb{C}^{n_{R,i,r}\times1}$
is to be transferred to RU $(i,r)$ on the fronthaul link, and $\mathbf{V}_{i,k}^{(i)}\in\mathbb{C}^{n_{R,i}\times d_{i,k,P}}$
is the precoding matrix for the signal $\mathbf{s}_{i,k,P}$.

\begin{figure}
\centering\includegraphics[width=13cm,height=5.5cm,keepaspectratio]{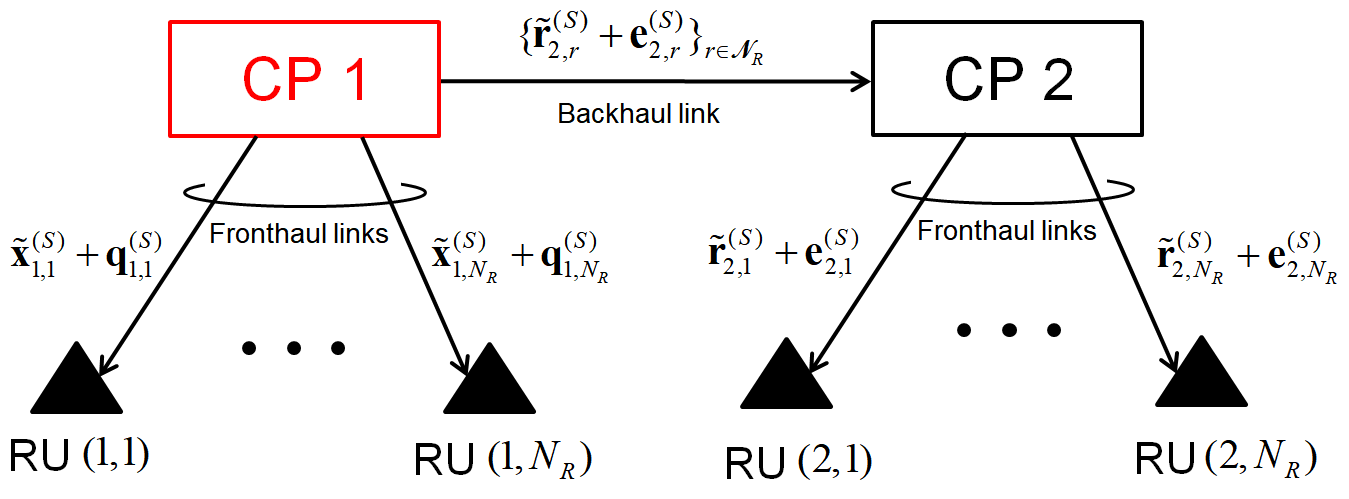}

\caption{\label{fig:FH-BH-quantization}Illustration of fronthaul and backhaul
quantization at CP 1.}
\end{figure}

\subsubsection{Linear Precoding for Shared Subband}

On the shared subband, the CPs and RUs of both operators are activated
to cooperatively serve all the UEs. To this end, CP $i$ precodes
the signal $\mathbf{s}_{i,k,S}$ for each UE $(i,k)$ into two precoded
signals: signal $\tilde{\mathbf{x}}_{i}^{(S)}\in\mathbb{C}^{n_{R,i}\times1}$
to be transmitted by its RUs and signal $\tilde{\mathbf{r}}_{\bar{i}}^{(S)}\in\mathbb{C}^{n_{R,\bar{i}}\times1}$
to be sent by the other operator $\bar{i}$. This is illustrated in
Fig. \ref{fig:FH-BH-quantization}. As we will discuss, the transmission
through the RUs of the other operator is enabled by the inter-CP backhaul
link and is subject to privacy constraints.

Mathematically, we write the precoded signals in the shared subband
as
\begin{align}
\tilde{\mathbf{x}}_{i}^{(S)}=\left[\tilde{\mathbf{x}}_{i,1}^{(S)\dagger}\,\cdots\,\tilde{\mathbf{x}}_{i,N_{R}}^{(S)\dagger}\right]^{\dagger}= & \sum_{k\in\mathcal{N}_{U}}\mathbf{V}_{i,k}^{(S)}\mathbf{s}_{i,k,S},\label{eq:precoding-shared-self}\\
\tilde{\mathbf{r}}_{\bar{i}}^{(S)}=\left[\tilde{\mathbf{r}}_{\bar{i},1}^{(S)\dagger}\,\cdots\,\tilde{\mathbf{r}}_{\bar{i},N_{R}}^{(S)\dagger}\right]^{\dagger}= & \sum_{k\in\mathcal{N}_{U}}\mathbf{T}_{\bar{i},k}^{(S)}\mathbf{s}_{i,k,S},\label{eq:precoding-shared-other}
\end{align}
where the subvectors $\tilde{\mathbf{x}}_{i,r}^{(S)}\in\mathbb{C}^{n_{R,i,r}\times1}$
and $\tilde{\mathbf{r}}_{\bar{i},r}^{(S)}\in\mathbb{C}^{n_{R,\bar{i},r}\times1}$
are communicated to the RUs $(i,r)$ and $(\bar{i},r)$, respectively;
and $\mathbf{V}_{i,k}^{(S)}\in\mathbb{C}^{n_{R,i}\times d_{i,k,S}}$
and $\mathbf{T}_{\bar{i},k}^{(S)}\in\mathbb{C}^{n_{R,\bar{i}}\times d_{i,k,S}}$
are the precoding matrices for the signal $\mathbf{s}_{i,k,S}$ associated
with the RUs $(i,r)$ and $(\bar{i},r)$, respectively.

\subsubsection{Fronthaul Compression}

CP $i$ is directly connected to the RUs $(i,r)$ in its network via
fronthaul links. Therefore, following the standard C-RAN operation,
the CP $i$ quantizes the precoded signals $\tilde{\mathbf{x}}_{i,r}^{(i)}$
and $\tilde{\mathbf{x}}_{i,r}^{(S)}$ for transmission on the fronthaul
link to RU $(i,r)$ on private and shared subbands. Assuming vector
quantization, we model the quantized signals $\hat{\mathbf{x}}_{i,r}^{(m)}$,
$m\in\{i,S\}$, as
\begin{equation}
\hat{\mathbf{x}}_{i,r}^{(m)}=\tilde{\mathbf{x}}_{i,r}^{(m)}+\mathbf{q}_{i,r}^{(m)},\label{eq:quantization-fronthaul}
\end{equation}
where $\mathbf{q}_{i,r}^{(m)}$ represents the quantization noise.
Adopting a Gaussian test channel as in \cite{Simeone-et-al:ETT}-\cite{Park-et-al:SPM},
the quantization noise $\mathbf{q}_{i,r}^{(m)}$ is independent of
the precoded signal $\tilde{\mathbf{x}}_{i,r}^{(m)}$ and distributed
as $\mathbf{q}_{i,r}^{(m)}\sim\mathcal{CN}(\mathbf{0},\mathbf{\Omega}_{i,r}^{(m)})$.
We recall that a Gaussian test channel can be well approximated by
vector lattice quantizers \cite{Ostergaard-Zamir}.

We first adopt standard point-to-point compression, whereby the signals
$\{\tilde{\mathbf{x}}_{i,r}^{(m)}\}_{r\in\mathcal{N}_{R},m\in\{i,S\}}$
for different RUs and subbands are compressed independently. A more
sophisticated approach based on multivariate compression will be discussed
in Sec. \ref{sec:Multivariate-Quantization}. Accordingly, with point-to-point
compression, the rate, in bit/s, needed to send $\hat{\mathbf{x}}_{i,r}^{(m)}$
to RU $(i,r)$ is given as $W_{i,m}I(\tilde{\mathbf{x}}_{i,r}^{(m)};\hat{\mathbf{x}}_{i,r}^{(m)})$
\cite[Ch. 3]{ElGamal-Kim}, where the mutual information $I(\tilde{\mathbf{x}}_{i,r}^{(m)};\hat{\mathbf{x}}_{i,r}^{(m)})$
can be written as
\begin{align}
I(\tilde{\mathbf{x}}_{i,r}^{(m)};\hat{\mathbf{x}}_{i,r}^{(m)}) & =g_{i,r}^{(m)}\left(\mathbf{V},\mathbf{\Omega}\right)\\
 & =\Phi\left(\sum_{k\in\mathcal{N}_{U}}\mathbf{K}\left(\mathbf{E}_{i,r}^{\dagger}\mathbf{V}_{i,k}^{(m)}\right),\,\mathbf{\Omega}_{i,r}^{(m)}\right).\nonumber
\end{align}
Here we defined the functions
\begin{equation}
\Phi(\mathbf{A},\mathbf{B})=\log_{2}\det(\mathbf{A}+\mathbf{B})-\log_{2}\det(\mathbf{B}),
\end{equation}
and $\mathbf{K}(\mathbf{A})=\mathbf{A}\mathbf{A}^{\dagger}$; the
shaping matrix $\mathbf{E}_{i,r}\in\mathbb{C}^{n_{R,i}\times n_{R,i,r}}$
that has all-zero elements except the rows from $\sum_{q=1}^{r-1}n_{R,i,q}+1$
to $\sum_{q=1}^{r}n_{R,i,q}$ which contains an identity matrix; and
the notation $W_{i,m}=W_{P,i}\cdot1(m=i)+W_{S}\cdot1(m=S)$.

\subsubsection{Backhaul Compression}

As seen in Fig. \ref{fig:FH-BH-quantization}, since there is no direct
link between CP $i$ and the RUs $(\bar{i},r)$ of the other tenant,
CP $i$ sends the precoded signal $\tilde{\mathbf{r}}_{\bar{i},r}^{(S)}$
to the RU $(\bar{i},r)$ through CP $\bar{i}$. The CP $\bar{i}$
forwards the received bit stream from CP $i$ to RU $(\bar{i},r)$.
Since both the backhaul link from CP $i$ to CP $\bar{i}$ and the
fronthaul link from CP $i$ to RU $(\bar{i},r)$ have finite capacities,
CP $i$ quantizes the signal $\tilde{\mathbf{r}}_{\bar{i},r}^{(S)}$
to obtain the quantized signal
\begin{equation}
\mathbf{r}_{\bar{i},r}^{(S)}=\tilde{\mathbf{r}}_{\bar{i},r}^{(S)}+\mathbf{e}_{\bar{i},r}^{(S)},\label{eq:quantization-backhaul}
\end{equation}
where $\mathbf{e}_{\bar{i},r}^{(S)}$ represents the quantization
noise. Using the same quantization model discussed above, this is
distributed as $\mathbf{e}_{\bar{i},r}^{(S)}\sim\mathcal{CN}(\mathbf{0},\mathbf{\Sigma}_{\bar{i},r}^{(S)})$.
As mentioned, we assume here the independent compression of the signals
$\{\tilde{\mathbf{r}}_{\bar{i},r}^{(S)}\}_{r\in\mathcal{N}_{R}}$
for different RUs, so that the rate needed to convey each signal $\mathbf{r}_{\bar{i},r}^{(S)}$
is given as $W_{S}I(\tilde{\mathbf{r}}_{\bar{i},r}^{(S)};\mathbf{r}_{\bar{i},r}^{(S)})$,
with
\begin{align}
I(\tilde{\mathbf{r}}_{\bar{i},r}^{(S)};\mathbf{r}_{\bar{i},r}^{(S)}) & =\gamma_{\bar{i},r}^{(S)}\left(\mathbf{T},\mathbf{\Sigma}\right)\\
 & =\Phi\left(\sum_{k\in\mathcal{N}_{U}}\mathbf{K}\left(\mathbf{E}_{\bar{i},r}^{\dagger}\mathbf{T}_{\bar{i},k}^{(S)}\right),\,\mathbf{\Sigma}_{\bar{i},r}^{(S)}\right).\nonumber
\end{align}
The capacity constraint for the backhaul link from CP $i$ to CP $\bar{i}$
can be written as
\begin{equation}
\sum_{r\in\mathcal{N}_{R}}W_{S}\gamma_{\bar{i},r}^{(S)}\left(\mathbf{T},\mathbf{\Sigma}\right)\leq C_{B,i},\,\,i\in\mathcal{N}_{O},\label{eq:backhaul-capacity-constraint}
\end{equation}
since the backhaul link needs to carry the baseband signals for all
the RUs. Similarly, the capacity constraint for the fronhtaul link
from CP $i$ to RU $(i,r)$ can be expressed as
\begin{align}
\sum_{m\in\{i,S\}}W_{i,m}g_{i,r}^{(m)}\left(\mathbf{V},\mathbf{\Omega}\right)+W_{S}\gamma_{i,r}^{(S)}\left(\mathbf{T},\mathbf{\Sigma}\right) & \leq C_{F,i,r},\,\,i\in\mathcal{N}_{O},\,r\in\mathcal{N}_{R},\label{eq:fronthaul-capacity-constraint}
\end{align}
since the fronthaul link needs to support transmission of the signals
for both private and shared subbands.

\subsubsection{Power Constraints}

The signals $\mathbf{x}_{i,r}^{(i)}$ and $\mathbf{x}_{i,r}^{(S)}$
transmitted by RU $(i,r)$ on the private and shared subbands are
given as $\mathbf{x}_{i,r}^{(i)}=\hat{\mathbf{x}}_{i,r}^{(i)}$ and
$\mathbf{x}_{i,r}^{(S)}=\hat{\mathbf{x}}_{i,r}^{(S)}+\mathbf{r}_{i,r}^{(S)}$,
respectively. We impose per-RU transmission power constraints as
\begin{align}
 & W_{P,i}p_{i,r}^{(i)}\left(\mathbf{V},\mathbf{\Omega}\right)+W_{S}p_{i,r}^{(S)}\left(\mathbf{V},\mathbf{T},\mathbf{\Omega}\right)\leq P_{i,r},\,\,i\in\mathcal{N}_{O},\,r\in\mathcal{N}_{R},\label{eq:power-RU-ir-total}
\end{align}
where $P_{i,r}$ represents the maximum transmission power allowed
for RU $(i,r)$, and the functions $p_{i,r}^{(i)}(\mathbf{V},\mathbf{\Omega},\mathbf{W})$
and $p_{i,r}^{(S)}(\mathbf{V},\mathbf{T},\mathbf{\Omega},\mathbf{W})$
measure the transmission powers per unit bandwidth on the private
and shared subbands, respectively, as
\begin{align}
p_{i,r}^{(i)}\left(\mathbf{V},\mathbf{\Omega},\mathbf{W}\right)\triangleq & \mathtt{E}\left\Vert \mathbf{x}_{i,r}^{(i)}\right\Vert ^{2}\label{eq:transmit-power-private}\\
= & \left(\sum_{k\in\mathcal{N}_{U}}\mathrm{tr}\left(\mathbf{K}\left(\mathbf{E}_{i,r}^{\dagger}\mathbf{V}_{i,k}^{(i)}\right)\right)+\mathrm{tr}\left(\mathbf{\Omega}_{i,r}^{(i)}\right)\right),\nonumber \\
p_{i,r}^{(S)}\left(\mathbf{V},\mathbf{T},\mathbf{\Omega},\mathbf{W}\right)\triangleq & \mathtt{E}\left\Vert \mathbf{x}_{i,r}^{(S)}\right\Vert ^{2}\label{eq:transmit-power-shared}\\
= & \left(\begin{array}{c}
\sum_{k\in\mathcal{N}_{U}}\mathrm{tr}\left(\mathbf{K}\left(\mathbf{E}_{i,r}^{\dagger}\mathbf{V}_{i,k}^{(S)}\right)\right)+\mathrm{tr}\left(\mathbf{\Omega}_{i,r}^{(S)}\right)\\
+\sum_{k\in\mathcal{N}_{U}}\mathrm{tr}\left(\mathbf{K}\left(\mathbf{E}_{i,r}^{\dagger}\mathbf{T}_{i,k}^{(S)}\right)\right)+\mathrm{tr}\left(\mathbf{\Sigma}_{i,r}^{(S)}\right)
\end{array}\right).\nonumber
\end{align}

\subsection{Decoding at UEs and Achievable Rates}

Each UE $(i,k)$ decodes the the submessage $M_{i,k,P}$ transmitted
on the private subband based on the received signal $\mathbf{y}_{i,k}^{(i)}$,
while treating the interference signals as additive noise. Then, the
maximum achievable rate $R_{i,k,P}$ of the submessage $M_{i,k,P}$
can be written as
\begin{align}
R_{i,k,P}= & W_{P,i}I(\mathbf{s}_{i,k,P};\,\mathbf{y}_{i,k}^{(i)}),\label{eq:achievable-rate-private}
\end{align}
where
\begin{align}
I(\mathbf{s}_{i,k,P};\,\mathbf{y}_{i,k}^{(i)}) & =f_{i,k,P}\left(\mathbf{V},\mathbf{\Omega}\right)\\
 & =\Phi\left(\mathbf{K}\left(\mathbf{H}_{i,k}^{i}\mathbf{V}_{i,k}^{(i)}\right),\,\sum_{l\in\mathcal{N}_{U}\setminus\{k\}}\mathbf{K}\left(\mathbf{H}_{i,k}^{i}\mathbf{V}_{i,l}^{(i)}\right)+\mathbf{H}_{i,k}^{i}\mathbf{\Omega}_{i}^{(i)}\mathbf{H}_{i,k}^{i\,\dagger}+\mathbf{I}\right).\nonumber
\end{align}
Here we defined the channel matrix $\mathbf{H}_{i,k}^{j}=[\mathbf{H}_{i,k}^{j,1}\,\mathbf{H}_{i,k}^{j,2}\,\cdots\,\mathbf{H}_{i,k}^{j,N_{R}}]$
from all the RUs of operator $j$ to UE $(i,k)$, and the matrix $\mathbf{\Omega}_{i}^{(i)}=\mathrm{diag}(\mathbf{\Omega}_{i,1}^{(i)},\ldots,\mathbf{\Omega}_{i,N_{R}}^{(i)})$.

In a similar manner, we assume that UE $(i,k)$ decodes the submessage
$M_{i,k,S}$ sent on the shared subband from the received signal $\mathbf{y}_{i,k}^{(S)}$
by treating the interference signals as noise, so that the maximum
achievable rate $R_{i,k,S}$ is given as
\begin{align}
R_{i,k,S}= & W_{S}I(\mathbf{s}_{i,k,S};\,\mathbf{y}_{i,k}^{(S)}),\label{eq:achievable-rate-shared}
\end{align}
with the mutual information $I(\mathbf{s}_{i,k,S};\,\mathbf{y}_{i,k}^{(S)})$
given as
\begin{align}
I(\mathbf{s}_{i,k,S};\,\mathbf{y}_{i,k}^{(S)}) & =f_{i,k,S}\left(\mathbf{V},\mathbf{T},\mathbf{\Omega}\right)\\
 & =\Phi\left(\mathbf{K}\left(\begin{array}{c}
\mathbf{H}_{i,k}^{i}\mathbf{V}_{i,k}^{(S)}\\
+\mathbf{H}_{i,k}^{\bar{i}}\mathbf{T}_{\bar{i},k}^{(S)}
\end{array}\right),\,\left(\begin{array}{c}
\sum_{l\in\mathcal{N}_{U}\setminus\{k\}}\mathbf{K}\left(\mathbf{H}_{i,k}^{i}\mathbf{V}_{i,l}^{(S)}+\mathbf{H}_{i,k}^{\bar{i}}\mathbf{T}_{\bar{i},l}^{(S)}\right)\\
+\sum_{l\in\mathcal{N}_{U}}\mathbf{K}\left(\mathbf{H}_{i,k}^{i}\mathbf{T}_{i,l}^{(S)}+\mathbf{H}_{i,k}^{\bar{i}}\mathbf{V}_{\bar{i},l}^{(S)}\right)\\
+\mathbf{H}_{i,k}^{i}\mathbf{\Omega}_{i}^{(S)}\mathbf{H}_{i,k}^{i\,\dagger}+\mathbf{H}_{i,k}^{i}\mathbf{\Sigma}_{i}^{(S)}\mathbf{H}_{i,k}^{i\,\dagger}\\
+\mathbf{H}_{i,k}^{\bar{i}}\mathbf{\Omega}_{\bar{i}}^{(S)}\mathbf{H}_{i,k}^{\bar{i}\,\dagger}+\mathbf{H}_{i,k}^{\bar{i}}\mathbf{\Sigma}_{\bar{i}}^{(S)}\mathbf{H}_{i,k}^{\bar{i}\,\dagger}+\mathbf{I}
\end{array}\right)\right),\nonumber
\end{align}
where we defined the matrices $\mathbf{\Omega}_{i}^{(S)}=\mathrm{diag}(\mathbf{\Omega}_{i,1}^{(S)},\ldots,\mathbf{\Omega}_{i,N_{R}}^{(S)})$
and $\mathbf{\Sigma}_{i}^{(S)}=\mathrm{diag}(\mathbf{\Sigma}_{i,1}^{(S)},\ldots,\mathbf{\Sigma}_{i,N_{R}}^{(S)})$.

\subsection{Privacy Constraints\label{sub:Privacy-Constraints}}

As discussed, inter-operator cooperation on the shared subband requires
the transmission of precoded and quantized signals $\{\mathbf{r}_{\bar{i},r}^{(S)}\}_{r\in\mathcal{N}_{R}}$
between CP $i$ and CP $\bar{i}$ on the backhaul link. As a result,
CP $\bar{i}$ can infer some information about the messages $\{M_{i,k,S}\}_{k\in\mathcal{N}_{U}}$
intended for the UEs $(i,k)$, $k\in\mathcal{N}_{U}$ of the operator
$i$. In order to ensure that this leakage of information is limited,
one can design both the precoding matrices $\mathbf{T}$ and the quantization
covariance matrices $\mathbf{\Sigma}$ under the information-theoretic
privacy constraint
\begin{align}
 & W_{S}I(\mathbf{s}_{i,k,S};\,\{\mathbf{r}_{\bar{i},r}^{(S)}\}_{r\in\mathcal{N}_{R}})\leq\Gamma_{\mathrm{privacy}}.\label{eq:privacy-constraint}
\end{align}
In (\ref{eq:privacy-constraint}), the mutual information $I(\mathbf{s}_{i,k,S};\,\{\mathbf{r}_{\bar{i},r}^{(S)}\}_{r\in\mathcal{N}_{R}})$
measures the amount of the information that can be inferred about
each signal $\mathbf{s}_{i,k,S}$ by the CP $\bar{i}$ of the other
operator based on the observation of $\{\mathbf{r}_{\bar{i},r}^{(S)}\}_{r\in\mathcal{N}_{R}}$.
This mutual information can be written as
\begin{align}
I(\mathbf{s}_{i,k,S};\,\{\mathbf{r}_{\bar{i},r}^{(S)}\}_{r\in\mathcal{N}_{R}}) & =\beta_{i,k,S}\left(\mathbf{T},\mathbf{\Omega}\right)\\
 & =\Phi\left(\mathbf{K}\left(\mathbf{T}_{\bar{i},k}^{(S)}\right),\,\sum_{l\in\mathcal{N}_{U}\setminus\{k\}}\mathbf{K}\left(\mathbf{T}_{\bar{i},l}^{(S)}\right)+\mathbf{\Sigma}_{\bar{i}}^{(S)}\right).\nonumber
\end{align}

The condition (\ref{eq:privacy-constraint}) imposes that the amount
of leaked information does not exceed a predetermined threshold value
$\Gamma_{\mathrm{privacy}}$. This value has a specific operational
meaning according to standard information-theoretic results \cite[Ch. 4, Problem 33]{Csiszar-Korner}.
In particular, a privacy level of $\Gamma_{\mathrm{privacy}}$ implies
that, if a user receives at rate $R$ (bit/s) on shared subband, a
bit stream of rate $\min(\Gamma_{\mathrm{privacy}},R)$ can be received
securely, while the remaining rate $(R-\Gamma_{\mathrm{privacy}})^{+}$
(bit/s) can be eavesdropped by the other operator.

In ensuring the satisfaction of the privacy constraint (\ref{eq:privacy-constraint}),
the quantization noise introduced by the fronthaul quantization plays
an important role. In fact, the fronthaul quantization noise is instrumental
in masking information about the UE messages at the cost of a more
significant degradation of the signals received by the UEs. A more
advanced quantization scheme will be considered in Sec. \ref{sec:Multivariate-Quantization}.

\section{Optimization of Multi-Tenant C-RAN\label{sec:Optimization}}

We aim at jointly optimizing the bandwidth allocation $\mathbf{W}$,
the precoding matrices $\{\mathbf{V},\mathbf{T}\}$ and the quantization
noise covariance matrices $\{\mathbf{\Omega},\mathbf{\Sigma}\}$,
with the goal of maximizing the sum-rate $R_{\Sigma}\triangleq\sum_{i\in\mathcal{N}_{O}}\sum_{k\in\mathcal{N}_{U}}(R_{i,k,P}+R_{i,k,S})$
of all the UEs, under constraints on backhaul and fronthaul capacity,
per-RU transmit power and inter-operator privacy levels. The problem
can be stated as\begin{subequations}\label{eq:problem-original}
\begin{align}
\underset{\mathbf{V},\mathbf{T},\boldsymbol{\Omega},\mathbf{\Sigma},\mathbf{W},\mathbf{R}}{\mathrm{maximize}}\, & \,\,\sum_{i\in\mathcal{N}_{O}}\sum_{k\in\mathcal{N}_{U}}(R_{i,k,P}+R_{i,k,S})\label{eq:problem-original-objective}\\
\mathrm{s.t.}\,\,\, & R_{i,k,P}\leq W_{P,i}f_{i,k,P}\left(\mathbf{V},\mathbf{\Omega}\right),\,\,i\in\mathcal{N}_{O},\,k\in\mathcal{N}_{U},\label{eq:problem-original-rate-P}\\
 & R_{i,k,S}\leq W_{S}f_{i,k,S}\left(\mathbf{V},\mathbf{T},\mathbf{\Omega}\right),\,\,i\in\mathcal{N}_{O},\,k\in\mathcal{N}_{U},\label{eq:problem-original-rate-S}\\
 & \sum_{r\in\mathcal{N}_{R}}W_{S}\gamma_{\bar{i},r}^{(S)}\left(\mathbf{T},\mathbf{\Sigma}\right)\leq C_{B,i},\,\,i\in\mathcal{N}_{O},\label{eq:problem-original-backhaul}\\
 & \sum_{m\in\{i,S\}}\!\!W_{i,m}g_{i,r}^{(m)}\left(\mathbf{V},\mathbf{\Omega}\right)\!+\!W_{S}\gamma_{i,r}^{(S)}\!\left(\mathbf{T},\mathbf{\Sigma}\right)\!\leq\!C_{F,i,r},\,i\in\mathcal{N}_{O},\,r\in\mathcal{N}_{R},\label{eq:problem-original-fronthaul}\\
 & W_{S}\beta_{i,k,S}\left(\mathbf{T},\mathbf{\Omega}\right)\leq\Gamma_{\mathrm{privacy}},\,\,i\in\mathcal{N}_{O},\,k\in\mathcal{N}_{U},\label{eq:problem-original-privacy}\\
 & W_{P,i}p_{i,r}^{(i)}\left(\mathbf{V},\mathbf{\Omega}\right)+W_{S}p_{i,r}^{(S)}\left(\mathbf{V},\mathbf{T},\mathbf{\Omega}\right)\leq P_{i,r},\,\,i\in\mathcal{N}_{O},\,r\in\mathcal{N}_{R},\label{eq:problem-original-power}\\
 & W_{P,1}+W_{P,2}+W_{S}=W.\label{eq:problem-original-BW}
\end{align}
\end{subequations}In (\ref{eq:problem-original}), constraints (\ref{eq:problem-original-rate-P})-(\ref{eq:problem-original-rate-S})
follow from the achievable rates (\ref{eq:achievable-rate-private})
and (\ref{eq:achievable-rate-shared}); (\ref{eq:problem-original-backhaul})-(\ref{eq:problem-original-fronthaul})
are the backhaul and fronthaul capacity constraints (\ref{eq:backhaul-capacity-constraint})
and (\ref{eq:fronthaul-capacity-constraint}); (\ref{eq:problem-original-privacy})
is the inter-operator privacy constraint (\ref{eq:privacy-constraint});
(\ref{eq:problem-original-power}) is the per-RU transmit power constraint
(\ref{eq:power-RU-ir-total}); and (\ref{eq:problem-original-BW})
is the sum-bandwidth constraint (\ref{eq:bandwidth-splitting}).

Since the problem (\ref{eq:problem-original}) is non-convex, we adopt
a Successive Convex Approximation (SCA) approach to obtain an efficient
local optimization algorithm. To this end, we equivalently rewrite
the constraints (\ref{eq:problem-original-rate-P}) and (\ref{eq:problem-original-rate-S})
using the epigraph form as
\begin{align}
 & \log R_{i,k,P}\leq\log W_{P,i}+\log t_{f,i,k,P},\,\,i\in\mathcal{N}_{O},\,k\in\mathcal{N}_{U},\label{eq:epigraph-rate-P-1}\\
 & t_{f,i,k,P}\leq f_{i,k,P}\left(\mathbf{V},\mathbf{\Omega}\right),\,\,i\in\mathcal{N}_{O},\,k\in\mathcal{N}_{U},\label{eq:epigraph-rate-P-2}\\
\mathrm{and}\,\, & \log R_{i,k,S}\leq\log W_{S}+\log t_{f,i,k,S},\,\,i\in\mathcal{N}_{O},\,k\in\mathcal{N}_{U},\label{eq:epigraph-rate-S-1}\\
 & t_{f,i,k,S}\leq f_{i,k,S}\left(\mathbf{V},\mathbf{T},\mathbf{\Omega}\right)\,\,i\in\mathcal{N}_{O},\,k\in\mathcal{N}_{U},\label{eq:epigraph-rate-S-2}
\end{align}
respectively. We note that the conditions (\ref{eq:epigraph-rate-P-1})
and (\ref{eq:epigraph-rate-S-1}) are difference-of-convex (DC) constraints
(see, e.g., \cite{Park-et-al:TSP13}\cite{Tao-et-al}), and that the
conditions (\ref{eq:epigraph-rate-P-2}) and (\ref{eq:epigraph-rate-S-2})
can be converted into DC constraints by expressing them with respect
to the variables $\tilde{\mathbf{V}}_{i,k}^{(i)}=\mathbf{V}_{i,k}^{(i)}\mathbf{V}_{i,k}^{(i)\dagger}$
and $\tilde{\mathbf{U}}_{i,k}^{(S)}=\mathbf{U}_{i,k}^{(S)}\mathbf{U}_{i,k}^{(S)\dagger}$
with $\mathbf{U}_{i,k}^{(S)}=[\mathbf{V}_{i,k}^{(S)\dagger}\,\mathbf{T}_{i,k}^{(S)\dagger}]^{\dagger}$.
Similarly, the other non-convex constraints (\ref{eq:problem-original-fronthaul})-(\ref{eq:problem-original-power})
can also be transformed into DC conditions by relaxing the non-convex
rank constraints $\mathrm{rank}(\tilde{\mathbf{V}}_{i,k}^{(i)})\leq d_{i,k,P}$
and $\mathrm{rank}(\tilde{\mathbf{U}}_{i,k}^{(S)})\leq d_{i,k,S}$.
As a result of these manipulations, we finally obtain the DC problem
(\ref{eq:problem-DC}) reported in Appendix \ref{appendix:DC-problem}.

We tackle the obtained DC problem by deriving an iterative algorithm
based on the standard CCCP approach \cite{Park-et-al:TSP13}\cite{Tao-et-al}.
The detailed algorithm is described in Algorithm 1. In our simulations,
we used the CVX software \cite{CVX} to solve the convex problem (\ref{eq:problem-convexified})
(see Appendix \ref{appendix:DC-problem}) at Step 2. After the convergence
of the algorithm, we need to project the variables $\tilde{\mathbf{V}}_{i,k}^{(i)\prime\prime}$
and $\tilde{\mathbf{U}}_{i,k}^{(S)\prime\prime}$ onto the spaces
of limited-rank matrices satisfying $\mathrm{rank}(\tilde{\mathbf{V}}_{i,k}^{(i)})\leq d_{i,k,P}$
and $\mathrm{rank}(\tilde{\mathbf{U}}_{i,k}^{(S)})\leq d_{i,k,S}$,
respectively. Without claim of optimality, we use the standard approach
of obtaining the variables $\tilde{\mathbf{V}}_{i,k}^{(i)}$ and $\tilde{\mathbf{U}}_{i,k}^{(S)}$
by including the $d_{i,k,P}$ and $d_{i,k,S}$ leading eigenvectors
of the matrices $\tilde{\mathbf{V}}_{i,k}^{(i)\prime\prime}$ and
$\tilde{\mathbf{U}}_{i,k}^{(S)\prime\prime}$, respectively, as columns.
\begin{algorithm}
\caption{CCCP algorithm for problem (\ref{eq:problem-DC})}

\textbf{1.} Initialize the variables $\tilde{\mathbf{V}}^{\prime}$,
$\tilde{\mathbf{U}}^{\prime}$, $\mathbf{\Omega}^{\prime}$, $\mathbf{\Sigma}^{\prime}$,
$\mathbf{W}^{\prime}$ and $\mathbf{R}^{\prime}$ to arbitrary feasible
values that satisfy the constraints (\ref{eq:problem-DC-rate-P-1})
and (\ref{eq:problem-DC-BW}) and set $q=1$.

\textbf{2.} Update the variables $\tilde{\mathbf{V}}^{\prime\prime}$,
$\tilde{\mathbf{U}}^{\prime\prime}$, $\mathbf{\Omega}^{\prime\prime}$,
$\mathbf{\Sigma}^{\prime\prime}$, $\mathbf{W}^{\prime\prime}$ and
$\mathbf{R}^{\prime\prime}$ as a solution of the convex problem (\ref{eq:problem-convexified})
in Appendix \ref{appendix:DC-problem}.

\textbf{3.} Stop if a convergence criterion is satisfied. Otherwise,
set $\tilde{\mathbf{V}}^{\prime}\leftarrow\tilde{\mathbf{V}}^{\prime\prime}$,
$\tilde{\mathbf{U}}^{\prime}\leftarrow\tilde{\mathbf{U}}^{\prime\prime}$,
$\mathbf{\Omega}^{\prime}\leftarrow\mathbf{\Omega}^{\prime\prime}$,
$\mathbf{\Sigma}^{\prime}\leftarrow\mathbf{\Sigma}^{\prime\prime}$,
$\mathbf{W}^{\prime}\leftarrow\mathbf{W}^{\prime\prime}$ and $\mathbf{R}^{\prime}\leftarrow\mathbf{R}^{\prime\prime}$
and go back to Step 2.
\end{algorithm}

\section{Multivariate Compression\label{sec:Multivariate-Quantization}}

In this section, we propose a novel quantization approach for inter-CP
communication that aims at controlling the trade-off between the distortion
at the UEs and inter-operator privacy. The approach is based on multivariate
compression, first studied for single-tenant systems in \cite{Park-et-al:TSP13}\cite{Lee-et-al:TSP16}.
To highlight the idea, we focus on the case of single RU per operator,
i.e., $N_{R}=1$, but extensions follow in the same way, albeit at
the cost of a more cumbersome notation.

The key idea is for each CP $i$ to jointly quantize the precoded
signals to be transmitted by the tenants' RUs. In so doing, one can
better control the impact of the quantization noise on the UEs' decoders,
while still ensuring a given level of privacy with respect to CP $\bar{i}$.

Mathematically, CP $i$ produces the linearly precoded signals $\tilde{\mathbf{x}}_{i,1}^{(S)}$
and $\tilde{\mathbf{r}}_{\bar{i},1}^{(S)}$ according to (\ref{eq:precoding-shared-self})
and (\ref{eq:precoding-shared-other}), respectively, and obtains
the quantized signals $\hat{\mathbf{x}}_{i,1}^{(S)}=\tilde{\mathbf{x}}_{i,1}^{(S)}+\mathbf{q}_{i,1}^{(S)}$
and $\mathbf{r}_{\bar{i},1}^{(S)}=\tilde{\mathbf{r}}_{\bar{i},1}^{(S)}+\mathbf{e}_{\bar{i},1}^{(S)}$
that are transferred to RUs $(i,1)$ and $(\bar{i},1)$, respectively.
With multivariate compression of the precoded signals $\tilde{\mathbf{x}}_{i,1}^{(S)}$
and $\tilde{\mathbf{r}}_{\bar{i},1}^{(S)}$, CP $i$ can ensure that
the quantization noise signals $\mathbf{q}_{i,1}^{(S)}$ and $\mathbf{e}_{\bar{i},1}^{(S)}$
have a correlation matrix $\mathbf{\Theta}_{i}^{(S)}=\mathtt{E}[\mathbf{q}_{i,1}^{(S)}\mathbf{e}_{\bar{i},1}^{(S)\dagger}]$.
As a result, the effective quantization noise signal that affects
the received signal $\mathbf{y}_{j,k}^{(S)}$ of UE $(j,k)$ on the
shared subband is given as $\tilde{\mathbf{q}}_{j,k}^{(S)}=\mathbf{H}_{j,k}^{i}\mathbf{q}_{i,1}^{(S)}+\mathbf{H}_{j,k}^{\bar{i}}\mathbf{e}_{\bar{i},1}^{(S)}$,
whose covariance matrix depends on the correlation matrix $\mathbf{\Theta}_{i}^{(S)}$
as
\begin{equation}
\mathtt{E}\left[\tilde{\mathbf{q}}_{j,k}^{(S)}\tilde{\mathbf{q}}_{j,k}^{(S)\dagger}\right]=\mathbf{G}_{j,k}^{i}\mathbf{\Lambda}_{i}\mathbf{G}_{j,k}^{i\,\dagger},\label{eq:effective-quantization-noise-covariance}
\end{equation}
where the matrix $\mathbf{\Lambda}_{i}$ represents the covariance
matrix of the stacked quantization noise signals $[\mathbf{q}_{i,1}^{(S)\dagger}\,\mathbf{e}_{\bar{i},1}^{(S)\dagger}]^{\dagger}$
as
\begin{equation}
\mathbf{\Lambda}_{i}=\mathtt{E}\left[\left[\begin{array}{c}
\mathbf{q}_{i,1}^{(S)}\\
\mathbf{e}_{\bar{i},1}^{(S)}
\end{array}\right]\left[\mathbf{q}_{i,1}^{(S)\dagger}\,\mathbf{e}_{\bar{i},1}^{(S)\dagger}\right]\right]=\left[\begin{array}{cc}
\mathbf{\Omega}_{i,1}^{(S)} & \mathbf{\Theta}_{i}^{(S)}\\
\mathbf{\Theta}_{i}^{(S)\dagger} & \mathbf{\Sigma}_{\bar{i},1}^{(S)}
\end{array}\right]\succeq\mathbf{0}.\label{eq:total-quantization-noise-covariance}
\end{equation}
Designing $\mathbf{\Theta}_{i}^{(S)}$ hence allows us to control
the effective noise observed by the UE, while also affecting the inter-operator
privacy constraint (\ref{eq:privacy-constraint}).

For the optimization under multivariate compression, it was shown
in \cite[Ch. 9]{ElGamal-Kim} that correlating the quantization noise
signals imposes the following additional constraint on the variables
$t_{g,i,1,S}$ and $t_{\gamma,\bar{i},r,S}$ in the DC problem (\ref{eq:problem-DC})
detailed in Appendix \ref{appendix:DC-problem}:
\begin{align}
 & h(\hat{\mathbf{x}}_{i,1}^{(S)})+h(\mathbf{r}_{\bar{i},1}^{(S)})-h(\hat{\mathbf{x}}_{i,1}^{(S)},\mathbf{r}_{\bar{i},1}^{(S)}|\tilde{\mathbf{x}}_{i,1}^{(S)},\tilde{\mathbf{r}}_{\bar{i},1}^{(S)})\nonumber \\
= & \log_{2}\det\left(\sum_{k\in\mathcal{N}_{U}}\tilde{\mathbf{E}}_{i,1}^{\dagger}\tilde{\mathbf{U}}_{i,k}^{(S)}\tilde{\mathbf{E}}_{i,1}+\mathbf{\Omega}_{i,1}^{(S)}\right)+\log_{2}\det\left(\sum_{k\in\mathcal{N}_{U}}\bar{\mathbf{E}}_{i,1}^{\dagger}\tilde{\mathbf{U}}_{i,k}^{(S)}\bar{\mathbf{E}}_{i,1}+\mathbf{\Sigma}_{\bar{i},1}^{(S)}\right)\nonumber \\
 & -\log_{2}\det\left(\mathbf{\Lambda}_{i}\right)\leq t_{g,i,1,S}+t_{\gamma,\bar{i},r,S}.\label{eq:multivariate-constraint}
\end{align}
The optimization under multivariate quantization is stated as the
problem (\ref{eq:problem-DC}) in Appendix \ref{appendix:DC-problem}
with the constraint (\ref{eq:multivariate-constraint}) added. We
can handle the problem following Algorithm 1 since the added condition
is a DC constraint.

\section{Numerical Results\label{sec:Numerical-Results}}

\begin{figure}
\centering\includegraphics[width=12cm,height=9.5cm]{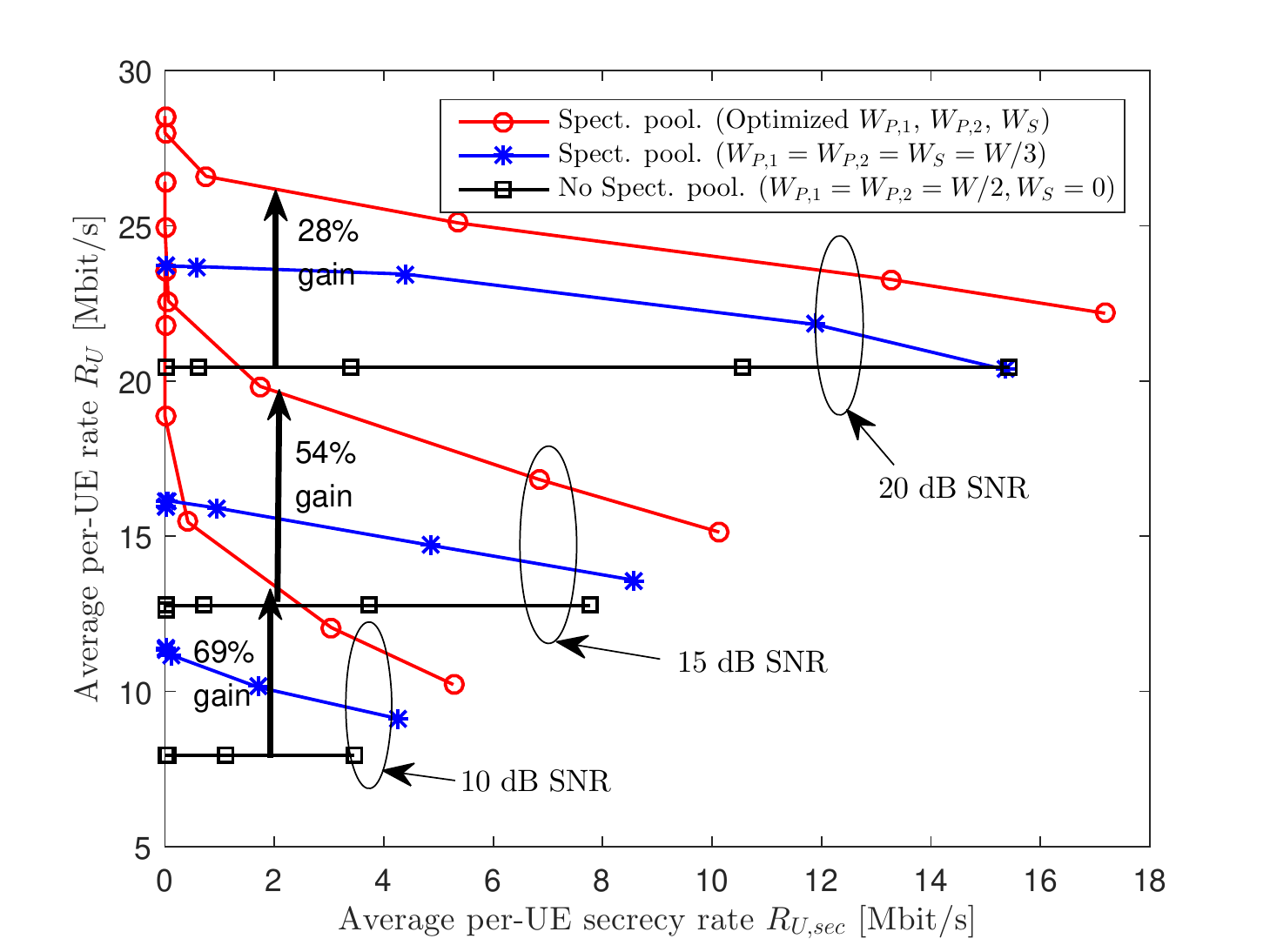}\caption{{\scriptsize{}\label{fig:graph-perUE-vs-perUEsec-SISO}}{\footnotesize{}Average
per-UE rate $R_{U}$ versus average per-UE secrecy rate $R_{U,\mathrm{sec}}$
($N_{R}=N_{U}=1$, $n_{R,i,r}=n_{U,i,k}=1$, $C_{B,i}=100$ Mbit/s,
$C_{F,i,r}=50$ Mbit/s and $W=10$ MHz).}}
\end{figure}

\begin{figure}
\centering\includegraphics[width=12cm,height=9.5cm]{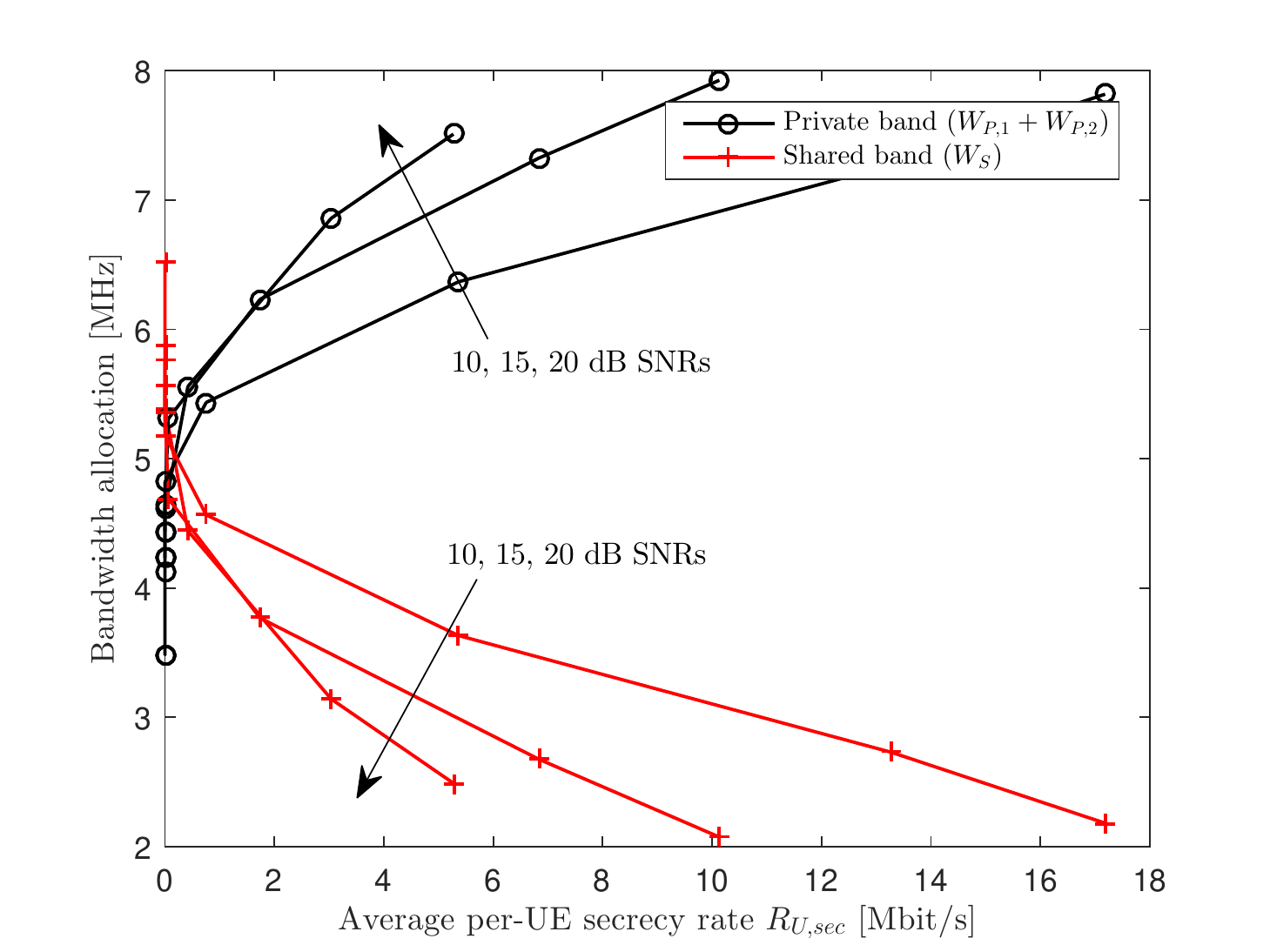}

\caption{{\scriptsize{}\label{fig:graph-BW-vs-perUEsec-SISO}}{\footnotesize{}Bandwidth
allocation versus average per-UE secrecy rate $R_{U,\mathrm{sec}}$
($N_{R}=N_{U}=1$, $n_{R,i,r}=n_{U,i,k}=1$, $C_{B,i}=100$ Mbit/s,
$C_{F,i,r}=50$ Mbit/s and $W=10$ MHz).}}
\end{figure}

\begin{figure}
\begin{minipage}[t]{0.45\columnwidth}%
\centering\includegraphics[width=9.1cm,height=7.2cm]{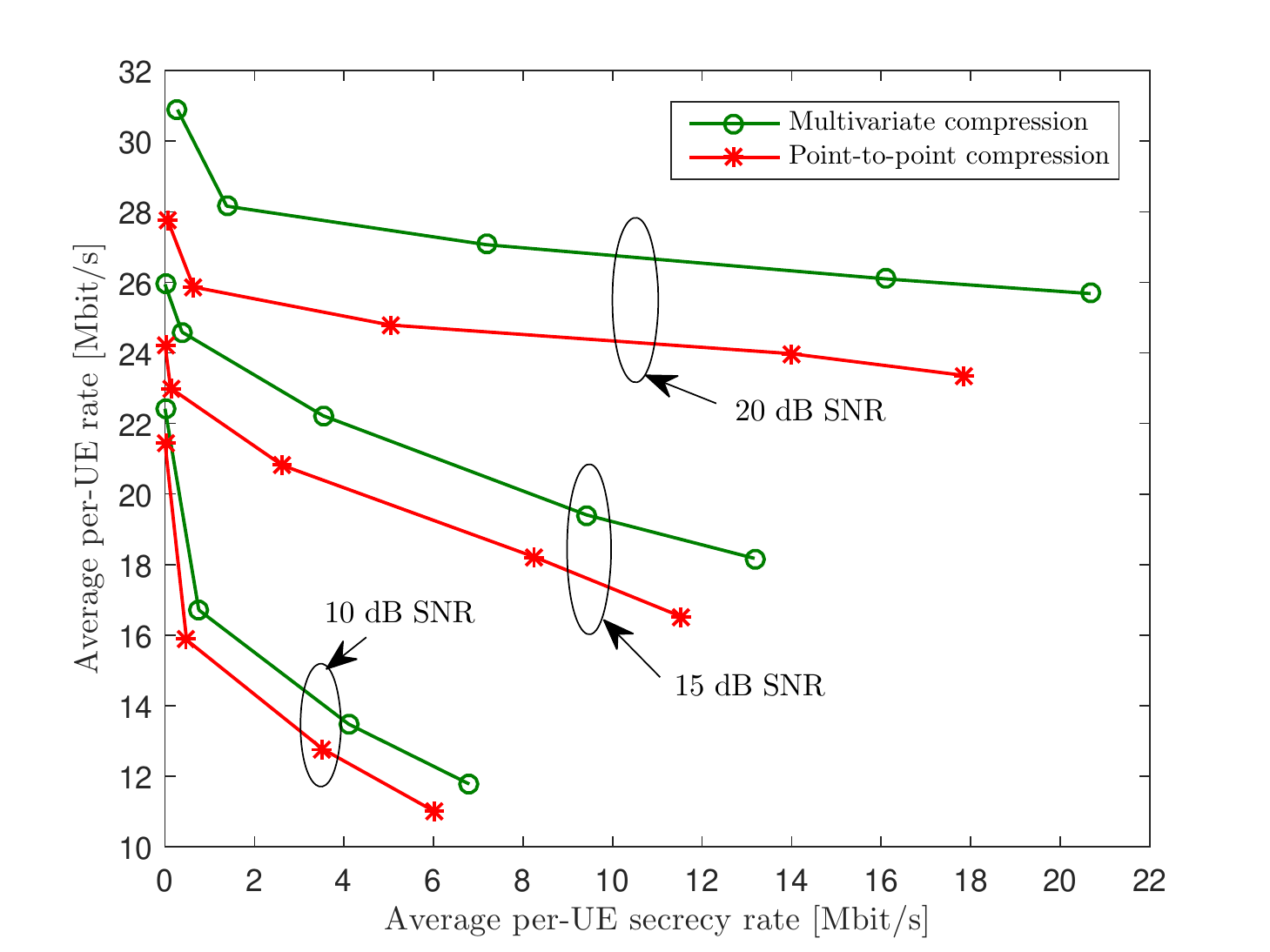}

\centering~~~~~{\footnotesize{}(a)}{\footnotesize \par}%
\end{minipage}\qquad{}%
\begin{minipage}[t]{0.45\columnwidth}%
\centering\includegraphics[width=9.1cm,height=7.2cm]{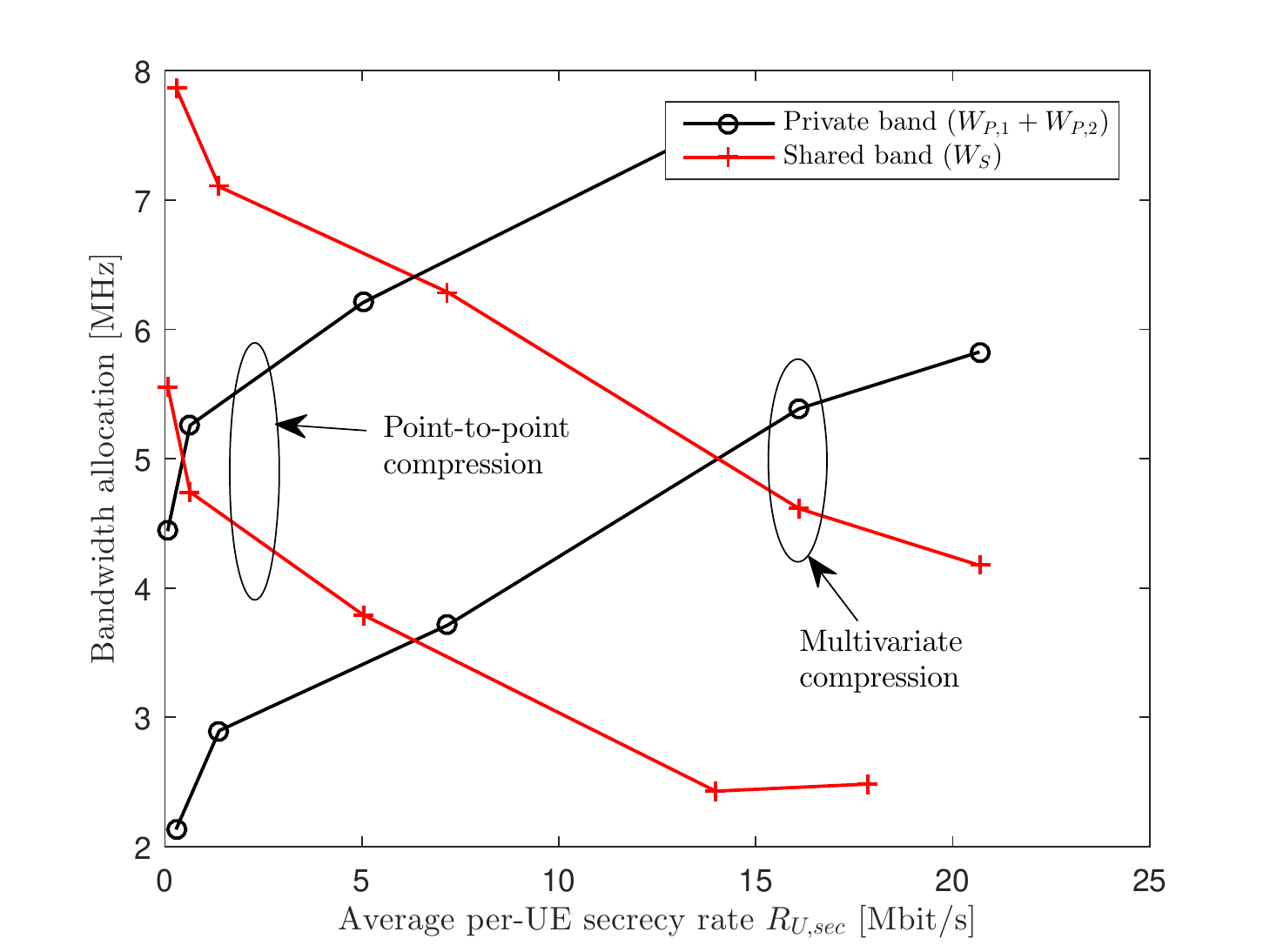}

\centering~~~~~{\footnotesize{}(b)}{\footnotesize \par}%
\end{minipage}\caption{{\scriptsize{}\label{fig:graph-perUE-vs-perUEsec-mulvar}}{\footnotesize{}(a)
Average per-UE rate $R_{U}$ and (b) bandwidth allocation versus average
per-UE secrecy rate $R_{U,\mathrm{sec}}$ ($N_{R}=N_{U}=1$, $n_{R,i,r}=n_{U,i,k}=1$,
$C_{B,i}=100$ Mbit/s, $C_{F,i,r}=50$ Mbit/s and $W=10$ MHz).}}
\end{figure}

In this section, we present numerical results that validate the performance
of multi-tenant C-RAN systems with spectrum pooling in the presence
of the proposed optimization and quantization strategies. We assume
that the positions of the RUs and UEs are uniformly distributed within
a circular area of radius 100 m. For given positions of the RUs and
the UEs, the channel matrix $\mathbf{H}_{i,k}^{j,r}$ from RU $(j,r)$
to UE $(i,k)$ is modeled as $\mathbf{H}_{i,k}^{j,r}=\sqrt{\rho_{i,k}^{j,r}}\tilde{\mathbf{H}}_{i,k}^{j,r}$,
where $\rho_{i,k}^{j,r}=1/(1+(D_{i,k}^{j,r}/D_{0})^{\alpha})$ represents
the path-loss, $D_{i,k}^{j,r}$ is the distance between the RU $(j,r)$
to UE $(i,k)$, and the elements of $\tilde{\mathbf{H}}_{i,k}^{j,r}$
are independent and identically distributed (i.i.d.) as $\mathcal{CN}(0,1)$.
In the simulation, we set $\alpha=3$ and $D_{0}=50$ m. Except for
Fig. \ref{fig:graph-perUE-vs-perUEsec-mulvar}, we focus on the point-to-point
compression strategy studied in Sec. \ref{sec:Multi-Tenant-C-RAN-With}
and Sec. \ref{sec:Optimization}.

To validate the effectiveness of the proposed designs, we compare
the following three schemes:

$\bullet$ \textit{Spectrum pooling with optimized bandwidth allocation
$W_{P,1}$, $W_{P,2}$ and $W_{S}$};

$\bullet$ \textit{Spectrum pooling with equal bandwidth allocation
$W_{P,1}=W_{P,2}=W_{S}=W/3$};

$\bullet$ \textit{No spectrum pooling with equal bandwidth allocation
$W_{P,1}=W_{P,2}=W/2$ and $W_{S}=0$}.

$\!\!\!\!\!\!$The first approach adopts the proposed optimization
algorithm (see Algorithm 1) discussed in Sec. \ref{sec:Optimization}.
Instead, the other two baseline approaches are obtained by using the
proposed algorithm with the added linear equality constraints $W_{P,1}=W_{P,2}=W_{S}=W/3$,
or $W_{P,1}=W_{P,2}=W/2$ and $W_{S}=0$, respectively. Except for
the last scheme with no spectrum pooling, all schemes exhibit a trade-off
between the achievable sum-rate and the guaranteed privacy level $\Gamma_{\mathrm{privacy}}$.
A smaller $\Gamma_{\mathrm{privacy}}$, i.e., a stricter privacy constraint,
generally entails a smaller sum-rate, and vice versa for a larger
$\Gamma_{\mathrm{privacy}}$. To quantify this effect, we define the
per-UE secrecy rate $R_{U,\mathrm{sec}}$ as $R_{U,\mathrm{sec}}=[R_{U}-\Gamma_{\mathrm{privacy}}]^{+}$,
where $R_{U}$ is the average per-UE achievable rate and $[\cdot]^{+}$
is defined as $[a]^{+}=\max\{a,0\}$. The rate $R_{U}$ is given by
the total sum-rate $R_{\Sigma}$ divided by the number of total UEs,
i.e., $R_{U}=R_{\Sigma}/(2N_{U})$. Following the discussion in Sec.
\ref{sub:Privacy-Constraints}, the quantity $R_{U,\mathrm{sec}}$
measures the rate at which information is transmitted privately to
each UE. In contrast, $R_{U}$ represents the overall transmission
rate, including both secure and insecure data streams.

Fig. \ref{fig:graph-perUE-vs-perUEsec-SISO} plots the average per-UE
rate $R_{U}$ versus the average per-UE secrecy rate $R_{U,\mathrm{sec}}$
for a multi-tenant C-RAN with $N_{R}=N_{U}=1$, $n_{R,i,r}=n_{U,i,k}=1$,
$C_{B,i}=100$ Mbit/s, $C_{F,i,r}=50$ Mbit/s, $W=10$ MHz and 10,
15 and 20 dB signal-to-noise ratios (SNRs). The curves are obtained
by varying the privacy threshold levels ranging from 5 Mbit/s to 60
Mbit/s in the constraints (\ref{eq:problem-original-privacy}). In
the figure, the multi-tenant C-RAN system with the proposed optimization
achieves a significantly improved rate-privacy trade-off as compared
to the other two strategies with no spectrum pooling or uniform spectrum
allocation. The gain becomes more significant at lower SNR levels,
since the impact of inter-operator cooperation in the shared subband
is more pronounced in this regime. As an example, in order to guarantee
the per-UE secrecy rate of 2 Mbit/s, the proposed multi-tenant C-RAN
system achieves a gain of about 69\% gain in terms of per-UE rates
at 10 dB SNR with respect to the traditional C-RAN system without
spectrum pooling.

Fig. \ref{fig:graph-BW-vs-perUEsec-SISO} plots the average bandwidth
allocation between the private and shared subbands versus the average
per-UE secrecy rate $R_{U,\mathrm{sec}}$ for the set-up considered
in Fig. \ref{fig:graph-perUE-vs-perUEsec-SISO}. Consistently with
the discussion above, as the SNR decreases, it is seen that more spectrum
resources are allocated to the shared subband in order to leverage
the opportunity of inter-operator cooperation.

\begin{figure}
\centering\includegraphics[width=12cm,height=9.5cm]{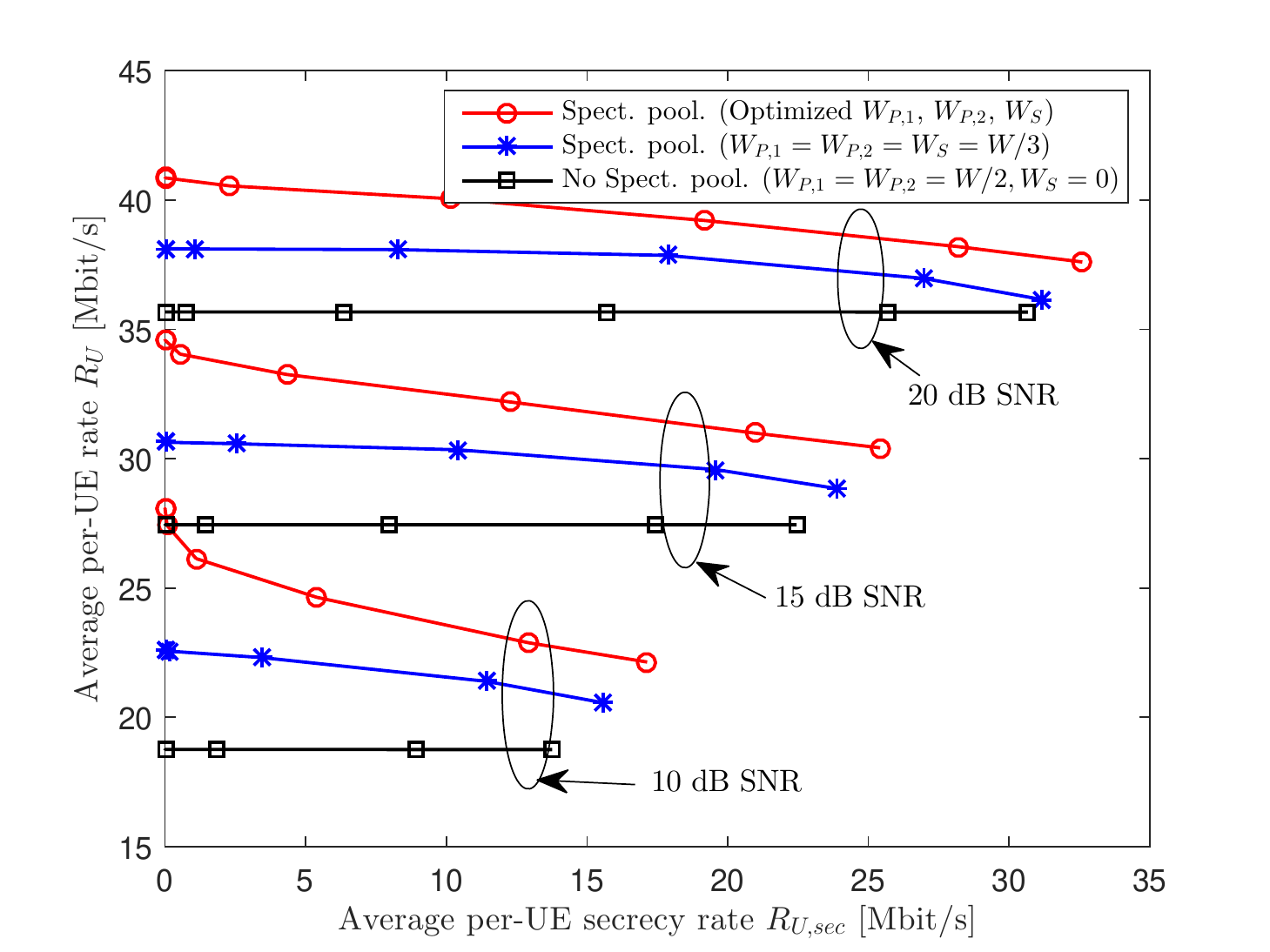}\caption{{\scriptsize{}\label{fig:graph-perUE-vs-perUEsec-2x2}}{\footnotesize{}Average
per-UE rate $R_{U}$ versus average per-UE secrecy rate $R_{U,\mathrm{sec}}$
($N_{R}=N_{U}=1$, $n_{R,i,r}=n_{U,i,k}=2$, $C_{B,i}=100$ Mbit/s,
$C_{F,i,r}=50$ Mbit/s and $W=10$ MHz).}}
\end{figure}

In Fig. \ref{fig:graph-perUE-vs-perUEsec-2x2}, we elaborate on the
effect of the number of antennas. To this end, we show the average
per-UE rate $R_{U}$ versus the average per-UE secrecy rate $R_{U,\mathrm{sec}}$
for a multi-tenant C-RAN with the same set-up discussed above except
for $n_{R,i,r}=n_{U,i,k}=2$ instead of $n_{R,i,r}=n_{U,i,k}=1$.
We can see that, compared to the single-antenna case, all the three
schemes become more robust to the privacy constraint with the increased
number of RU and UE antennas. This is due to the additional degrees
of freedom in the precoding design that is afforded by the larger
number of antennas.

We now study the impact of correlating the quantization noise signals
across the RUs of operators by means of the multivariate compression
strategy proposed in Sec. \ref{sec:Multivariate-Quantization}. In
Fig. \ref{fig:graph-perUE-vs-perUEsec-mulvar}(a), we plot the average
per-UE rate $R_{U}$ versus the average per-UE secrecy rate $R_{U,\mathrm{sec}}$
for the same multi-tenant C-RAN set-up considered in Fig. \ref{fig:graph-perUE-vs-perUEsec-SISO}
assuming spectrum pooling with optimized bandwidths $\{W_{P,1},W_{P,2},W_{S}\}$.
We observe that multivariate compression is instrumental in improving
the trade-off between inter-operator cooperation and privacy. The
accrued performance gain increases with the SNR, since the performance
degradation due to quantization is masked by the additive noise when
the SNR is small. Fig. \ref{fig:graph-perUE-vs-perUEsec-mulvar}(b)
plots the optimized bandwidth allocation versus the average per-UE
secrecy rate $R_{U,\mathrm{sec}}$ for a 20 dB SNR. The figure suggests
that, with multivariate compression, it is desirable to allocate more
bandwidth to the shared subband, given the added benefits of inter-operator
cooperation in the presence of multivariate compression.

\section{Conclusions\label{sec:Conclusion} }

In this work, we have studied the design of multi-tenant C-RAN systems
with spectrum pooling under inter-operator privacy constraints. Assuming
the standard C-RAN operation with quantized baseband signals, we first
considered the standard point-to-point compression strategy, and then
proposed a novel multivariate compression to achieve a better trade-off
between the inter-operator cooperation and privacy. For both cases,
we tackled the joint optimization of the bandwidth allocation among
the private and shared subbands and of the precoding and fronthaul/backhaul
compression strategies while satisfying constraints on fronthaul and
backhaul capacity, per-RU transmit power and inter-operator privacy
levels. To tackle the non-convex optimization problems, we converted
the problems into DC problems with rank relaxation and derived iterative
algorithms based on the standard CCCP. We provided extensive numerical
results to validate the effectiveness of the multi-tenant C-RAN system
with the proposed optimization algorithm and multivariate compression.
Among open problems, we mention the extension of the design and analysis
to models with RAN sharing at the level of RUs and the consideration
of hierarchical fog architectures.

\appendices

\section{}\label{appendix:DC-problem}

By relaxing the non-convex rank constraints $\mathrm{rank}(\tilde{\mathbf{V}}_{i,k}^{(i)})\leq d_{i,k,P}$
and $\mathrm{rank}(\tilde{\mathbf{U}}_{i,k}^{(S)})\leq d_{i,k,S}$
explained in Sec. \ref{sec:Optimization}, the problem (\ref{eq:problem-original})
can be converted into the DC problem\begin{subequations}\label{eq:problem-DC}
\begin{align}
\underset{_{\mathbf{R},\mathbf{t},\tilde{\mathbf{g}},\tilde{\boldsymbol{\gamma}},\tilde{\mathbf{p}}}^{\tilde{\mathbf{V}},\tilde{\mathbf{U}},\boldsymbol{\Omega},\mathbf{\Sigma},\mathbf{W},}}{\mathrm{maximize}}\, & \,\,\sum_{i\in\mathcal{N}_{O}}\sum_{k\in\mathcal{N}_{U}}(R_{i,k,P}+R_{i,k,S})\label{eq:problem-DC-objective}\\
\mathrm{s.t.}\,\,\, & \log R_{i,k,P}\leq\log W_{P,i}+\log t_{f,i,k,P},\,\,i\in\mathcal{N}_{O},\,k\in\mathcal{N}_{U},\label{eq:problem-DC-rate-P-1}\\
 & t_{f,i,k,P}\leq f_{i,k,P}\left(\tilde{\mathbf{V}},\mathbf{\Omega}\right),\,\,i\in\mathcal{N}_{O},\,k\in\mathcal{N}_{U},\label{eq:problem-DC-rate-P-2}\\
 & \log R_{i,k,S}\leq\log W_{S}+\log t_{f,i,k,S},\,\,i\in\mathcal{N}_{O},\,k\in\mathcal{N}_{U},\label{eq:problem-DC-rate-S-1}\\
 & t_{f,i,k,S}\leq f_{i,k,S}\left(\tilde{\mathbf{V}},\tilde{\mathbf{U}},\mathbf{\Omega}\right)\,\,i\in\mathcal{N}_{O},\,k\in\mathcal{N}_{U},\label{eq:problem-DC-rate-S-2}\\
 & \sum_{m\in\{i,S\}}\tilde{g}_{i,r}^{(m)}+\tilde{\gamma}_{i,r}^{(S)}\leq C_{F,i,r},\,\,i\in\mathcal{N}_{O},\,r\in\mathcal{N}_{R},\label{eq:problem-DC-fronthaul-1}\\
 & \log W_{i,m}+\log t_{g,i,r,m}\leq\log\tilde{g}_{i,r}^{(m)},\,i\in\mathcal{N}_{O},\,r\in\mathcal{N}_{R},\,m\in\{i,S\},\label{eq:problem-DC-fronthaul-2}\\
 & g_{i,r}^{(i)}\left(\tilde{\mathbf{V}},\mathbf{\Omega}\right)\leq t_{g,i,r,i},\,\,i\in\mathcal{N}_{O},\,r\in\mathcal{N}_{R},\label{eq:problem-DC-fronthaul-3}\\
 & g_{i,r}^{(S)}\left(\tilde{\mathbf{U}},\mathbf{\Omega}\right)\leq t_{g,i,r,S},\,\,i\in\mathcal{N}_{O},\,r\in\mathcal{N}_{R},\label{eq:problem-DC-fronthaul-4}\\
 & \log W_{S}+\log t_{\gamma,i,r,S}\leq\log\tilde{\gamma}_{i,r}^{(S)},\,\,i\in\mathcal{N}_{O},\,r\in\mathcal{N}_{R},\label{eq:problem-DC-fronhtaul-5}\\
 & \gamma_{i,r}^{(S)}\left(\tilde{\mathbf{U}},\mathbf{\Sigma}\right)\leq t_{\gamma,i,r,S},\,\,i\in\mathcal{N}_{O},\,r\in\mathcal{N}_{R},\label{eq:problem-DC-fronthaul-6}\\
 & \sum_{r\in\mathcal{N}_{R}}\tilde{\gamma}_{\bar{i},r}^{(S)}\leq C_{B,i},\,\,i\in\mathcal{N}_{O},\label{eq:problem-DC-backhaul}\\
 & \log W_{S}+\log t_{\beta,i,k,S}\leq\log\Gamma_{\mathrm{privacy}},\,\,i\in\mathcal{N}_{O},\,k\in\mathcal{N}_{U},\label{eq:problem-DC-privacy-1}\\
 & \beta_{i,k,S}\left(\tilde{\mathbf{U}},\mathbf{\Omega}\right)\leq t_{\beta,i,k,S},\,\,i\in\mathcal{N}_{O},\,k\in\mathcal{N}_{U},\label{eq:problem-DC-privacy-2}\\
 & \tilde{p}_{i,r}^{(i)}+\tilde{p}_{i,r}^{(S)}\leq P_{i,r},\,\,i\in\mathcal{N}_{O},\,r\in\mathcal{N}_{R},\label{eq:problem-DC-power-1}\\
 & \log W_{i,m}+\log t_{p,i,r,m}\leq\log\tilde{p}_{i,r}^{(m)},\,\,i\in\mathcal{N}_{O},\,r\in\mathcal{N}_{R},\,m\in\{i,S\},\label{eq:problem-DC-power-2}\\
 & p_{i,r}^{(m)}\left(\tilde{\mathbf{V}},\tilde{\mathbf{U}},\mathbf{\Omega}\right)\leq t_{p,i,r,m},\,\,i\in\mathcal{N}_{O},\,r\in\mathcal{N}_{R},\,m\in\{i,S\},\label{eq:problem-DC-power-3}\\
 & W_{P,1}+W_{P,2}+W_{S}=W.\label{eq:problem-DC-BW}
\end{align}
\end{subequations}

Furthermore, at Step 2 in Algorithm 1, the CCCP approach solves the
convex problem obtained by linearizing the terms that induce non-convexity
of problem (\ref{eq:problem-DC}). This can be written as\begin{subequations}\label{eq:problem-convexified}
\begin{align}
\underset{_{\mathbf{R},\mathbf{t},\tilde{\mathbf{g}},\tilde{\boldsymbol{\gamma}},\tilde{\mathbf{p}}}^{\tilde{\mathbf{V}},\tilde{\mathbf{U}},\boldsymbol{\Omega},\mathbf{\Sigma},\mathbf{W},}}{\mathrm{maximize}}\, & \,\,\sum_{i\in\mathcal{N}_{O}}\sum_{k\in\mathcal{N}_{U}}(R_{i,k,P}+R_{i,k,S})\label{eq:problem-convexified-objective}\\
\mathrm{s.t.}\,\,\, & \varphi\left(R_{i,k,P},R_{i,k,P}^{\prime}\right)\leq\log W_{P,i}+\log t_{f,i,k,P},\,\,i\in\mathcal{N}_{O},\,k\in\mathcal{N}_{U},\label{eq:problem-convexified-rate-P-1}\\
 & t_{f,i,k,P}\leq\hat{f}_{i,k,P}\left(\tilde{\mathbf{V}},\mathbf{\Omega},\tilde{\mathbf{V}}^{\prime},\mathbf{\Omega}^{\prime}\right),\,\,i\in\mathcal{N}_{O},\,k\in\mathcal{N}_{U},\label{eq:problem-convexified-rate-P-2}\\
 & \varphi\left(R_{i,k,S},R_{i,k,S}^{\prime}\right)\leq\log W_{S}+\log t_{f,i,k,S},\,\,i\in\mathcal{N}_{O},\,k\in\mathcal{N}_{U},\label{eq:problem-convexified-rate-S-1}\\
 & t_{f,i,k,S}\leq\hat{f}_{i,k,S}\left(\tilde{\mathbf{V}},\tilde{\mathbf{U}},\mathbf{\Omega},\tilde{\mathbf{V}}^{\prime},\tilde{\mathbf{U}}^{\prime},\mathbf{\Omega}^{\prime}\right)\,\,i\in\mathcal{N}_{O},\,k\in\mathcal{N}_{U},\label{eq:problem-convexified-rate-S-2}\\
 & \sum_{m\in\{i,S\}}\tilde{g}_{i,r}^{(m)}+\tilde{\gamma}_{i,r}^{(S)}\leq C_{F,i,r},\,\,i\in\mathcal{N}_{O},\,r\in\mathcal{N}_{R},\label{eq:problem-convexified-fronthaul-1}\\
 & \varphi\left(W_{i,m},W_{i,m}^{\prime}\!\right)\!+\!\varphi\!\left(t_{g,i,r,m},t_{g,i,r,m}^{\prime}\!\right)\!\leq\!\log\tilde{g}_{i,r}^{(m)},\,i\in\mathcal{N}_{O},\,r\in\mathcal{N}_{R},\,m\in\{i,S\},\label{eq:problem-convexified-fronthaul-2}\\
 & \hat{g}_{i,r}^{(i)}\left(\tilde{\mathbf{V}},\mathbf{\Omega},\tilde{\mathbf{V}}^{\prime},\mathbf{\Omega}^{\prime}\right)\leq t_{g,i,r,i},\,\,i\in\mathcal{N}_{O},\,r\in\mathcal{N}_{R},\label{eq:problem-convexified-fronthaul-3}\\
 & \hat{g}_{i,r}^{(S)}\left(\tilde{\mathbf{U}},\mathbf{\Omega},\tilde{\mathbf{U}}^{\prime},\mathbf{\Omega}^{\prime}\right)\leq t_{g,i,r,S},\,\,i\in\mathcal{N}_{O},\,r\in\mathcal{N}_{R},\label{eq:problem-convexified-fronthaul-4}\\
 & \varphi\left(W_{S},W_{S}^{\prime}\right)+\varphi\left(t_{\gamma,i,r,S},t_{\gamma,i,r,S}^{\prime}\right)\leq\log\tilde{\gamma}_{i,r}^{(S)},\,\,i\in\mathcal{N}_{O},\,r\in\mathcal{N}_{R},\label{eq:problem-convexified-fronhtaul-5}\\
 & \hat{\gamma}_{i,r}^{(S)}\left(\tilde{\mathbf{U}},\mathbf{\Sigma},\tilde{\mathbf{U}}^{\prime},\mathbf{\Sigma}^{\prime}\right)\leq t_{\gamma,i,r,S},\,\,i\in\mathcal{N}_{O},\,r\in\mathcal{N}_{R},\label{eq:problem-convexified-fronthaul-6}\\
 & \sum_{r\in\mathcal{N}_{R}}\tilde{\gamma}_{\bar{i},r}^{(S)}\leq C_{B,i},\,\,i\in\mathcal{N}_{O},\label{eq:problem-convexified-backhaul}\\
 & \varphi\left(W_{S},W_{S}^{\prime}\right)+\varphi\left(t_{\beta,i,k,S},t_{\beta,i,k,S}^{\prime}\right)\leq\log\Gamma_{\mathrm{privacy}},\,\,i\in\mathcal{N}_{O},\,k\in\mathcal{N}_{U},\label{eq:problem-convexified-privacy-1}\\
 & \hat{\beta}_{i,k,S}\left(\tilde{\mathbf{U}},\mathbf{\Omega},\tilde{\mathbf{U}}^{\prime},\mathbf{\Omega}^{\prime}\right)\leq t_{\beta,i,k,S},\,\,i\in\mathcal{N}_{O},\,k\in\mathcal{N}_{U},\label{eq:problem-convexified-privacy-2}\\
 & \tilde{p}_{i,r}^{(i)}+\tilde{p}_{i,r}^{(S)}\leq P_{i,r},\,\,i\in\mathcal{N}_{O},\,r\in\mathcal{N}_{R},\label{eq:problem-convexified-power-1}\\
 & \varphi\left(\!W_{i,m},W_{i,m}^{\prime}\!\right)\!+\!\varphi\left(\!t_{p,i,r,m},t_{p,i,r,m}^{\prime}\!\right)\!\leq\!\log\tilde{p}_{i,r}^{(m)},\,\,i\in\mathcal{N}_{O},\,r\in\mathcal{N}_{R},\,m\in\{i,S\},\label{eq:problem-convexified-power-2}\\
 & p_{i,r}^{(m)}\left(\tilde{\mathbf{V}},\tilde{\mathbf{U}},\mathbf{\Omega}\right)\leq t_{p,i,r,m},\,\,i\in\mathcal{N}_{O},\,r\in\mathcal{N}_{R},\,m\in\{i,S\},\label{eq:problem-convexified-power-3}\\
 & W_{P,1}+W_{P,2}+W_{S}=W,\label{eq:problem-convexified-BW}
\end{align}
\end{subequations}where we defined the functions
\begin{align}
\hat{f}_{i,k,P}\left(\tilde{\mathbf{V}},\mathbf{\Omega},\tilde{\mathbf{V}}^{\prime},\mathbf{\Omega}^{\prime}\right)= & \log_{2}\det\left(\sum_{l\in\mathcal{N}_{U}}\mathbf{H}_{i,k}^{i}\tilde{\mathbf{V}}_{i,l}^{(i)}\mathbf{H}_{i,k}^{i\dagger}+\mathbf{H}_{i,k}^{i}\mathbf{\Omega}_{i}^{(i)}\mathbf{H}_{i,k}^{i\,\dagger}+\mathbf{I}\right)\\
 & -\frac{1}{\ln2}\varphi\left(\begin{array}{c}
\sum_{l\in\mathcal{N}_{U}\setminus\{k\}}\mathbf{H}_{i,k}^{i}\tilde{\mathbf{V}}_{i,l}^{(i)}\mathbf{H}_{i,k}^{i\dagger}+\mathbf{H}_{i,k}^{i}\mathbf{\Omega}_{i}^{(i)}\mathbf{H}_{i,k}^{i\,\dagger}+\mathbf{I},\\
\sum_{l\in\mathcal{N}_{U}\setminus\{k\}}\mathbf{H}_{i,k}^{i}\tilde{\mathbf{V}}_{i,l}^{(i)\prime}\mathbf{H}_{i,k}^{i\dagger}+\mathbf{H}_{i,k}^{i}\mathbf{\Omega}_{i}^{(i)\,\prime}\mathbf{H}_{i,k}^{i\,\dagger}+\mathbf{I}
\end{array}\right),\nonumber \\
\hat{f}_{i,k,S}\left(\tilde{\mathbf{V}},\tilde{\mathbf{U}},\mathbf{\Omega},\tilde{\mathbf{V}}^{\prime},\tilde{\mathbf{U}}^{\prime},\mathbf{\Omega}^{\prime}\right)= & \log_{2}\det\left(\begin{array}{c}
\sum_{l\in\mathcal{N}_{U}}\mathbf{G}_{i,k}^{i}\tilde{\mathbf{U}}_{i,l}^{(S)}\mathbf{G}_{i,k}^{i\,\dagger}\\
+\sum_{l\in\mathcal{N}_{U}}\mathbf{G}_{i,k}^{\bar{i}}\tilde{\mathbf{U}}_{\bar{i},l}^{(S)}\mathbf{G}_{i,k}^{\bar{i}\,\dagger}\\
+\mathbf{H}_{i,k}^{i}\mathbf{\Omega}_{i}^{(S)}\mathbf{H}_{i,k}^{i\,\dagger}+\mathbf{H}_{i,k}^{i}\mathbf{\Sigma}_{i}^{(S)}\mathbf{H}_{i,k}^{i\,\dagger}\\
+\mathbf{H}_{i,k}^{\bar{i}}\mathbf{\Omega}_{\bar{i}}^{(S)}\mathbf{H}_{i,k}^{\bar{i}\,\dagger}+\mathbf{H}_{i,k}^{\bar{i}}\mathbf{\Sigma}_{\bar{i}}^{(S)}\mathbf{H}_{i,k}^{\bar{i}\,\dagger}+\mathbf{I}
\end{array}\right)\\
 & -\frac{1}{\ln2}\varphi\left(\begin{array}{c}
\left(\begin{array}{c}
\sum_{l\in\mathcal{N}_{U}\setminus\{k\}}\mathbf{G}_{i,k}^{i}\tilde{\mathbf{U}}_{i,l}^{(S)}\mathbf{G}_{i,k}^{i\,\dagger}\\
+\sum_{l\in\mathcal{N}_{U}}\mathbf{G}_{i,k}^{\bar{i}}\tilde{\mathbf{U}}_{\bar{i},l}^{(S)}\mathbf{G}_{i,k}^{\bar{i}\,\dagger}\\
+\mathbf{H}_{i,k}^{i}\mathbf{\Omega}_{i}^{(S)}\mathbf{H}_{i,k}^{i\,\dagger}+\mathbf{H}_{i,k}^{i}\mathbf{\Sigma}_{i}^{(S)}\mathbf{H}_{i,k}^{i\,\dagger}\\
+\mathbf{H}_{i,k}^{\bar{i}}\mathbf{\Omega}_{\bar{i}}^{(S)}\mathbf{H}_{i,k}^{\bar{i}\,\dagger}+\mathbf{H}_{i,k}^{\bar{i}}\mathbf{\Sigma}_{\bar{i}}^{(S)}\mathbf{H}_{i,k}^{\bar{i}\,\dagger}+\mathbf{I}
\end{array}\right),\\
\left(\begin{array}{c}
\sum_{l\in\mathcal{N}_{U}\setminus\{k\}}\mathbf{G}_{i,k}^{i}\tilde{\mathbf{U}}_{i,l}^{(S)\prime}\mathbf{G}_{i,k}^{i\,\dagger}\\
+\sum_{l\in\mathcal{N}_{U}}\mathbf{G}_{i,k}^{\bar{i}}\tilde{\mathbf{U}}_{\bar{i},l}^{(S)\prime}\mathbf{G}_{i,k}^{\bar{i}\,\dagger}\\
+\mathbf{H}_{i,k}^{i}\mathbf{\Omega}_{i}^{(S)\,\prime}\mathbf{H}_{i,k}^{i\,\dagger}+\mathbf{H}_{i,k}^{i}\mathbf{\Sigma}_{i}^{(S)\,\prime}\mathbf{H}_{i,k}^{i\,\dagger}\\
+\mathbf{H}_{i,k}^{\bar{i}}\mathbf{\Omega}_{\bar{i}}^{(S)\,\prime}\mathbf{H}_{i,k}^{\bar{i}\,\dagger}+\mathbf{H}_{i,k}^{\bar{i}}\mathbf{\Sigma}_{\bar{i}}^{(S)\,\prime}\mathbf{H}_{i,k}^{\bar{i}\,\dagger}+\mathbf{I}
\end{array}\right)
\end{array}\right),\nonumber \\
\hat{g}_{i,r}^{(i)}\left(\tilde{\mathbf{V}},\mathbf{\Omega},\tilde{\mathbf{V}}^{\prime},\mathbf{\Omega}^{\prime}\right)= & \frac{1}{\ln2}\varphi\left(\sum_{k\in\mathcal{N}_{U}}\mathbf{E}_{i,r}^{\dagger}\tilde{\mathbf{V}}_{i,k}^{(i)}\mathbf{E}_{i,r}+\mathbf{\Omega}_{i,r}^{(i)},\,\sum_{k\in\mathcal{N}_{U}}\mathbf{E}_{i,r}^{\dagger}\tilde{\mathbf{V}}_{i,k}^{(i)\prime}\mathbf{E}_{i,r}+\mathbf{\Omega}_{i,r}^{(i)\,\prime}\right)\nonumber \\
 & -\log_{2}\det\left(\mathbf{\Omega}_{i,r}^{(i)}\right),\\
\hat{g}_{i,r}^{(S)}\left(\tilde{\mathbf{U}},\mathbf{\Omega},\tilde{\mathbf{U}}^{\prime},\mathbf{\Omega}^{\prime}\right)= & \frac{1}{\ln2}\varphi\left(\sum_{k\in\mathcal{N}_{U}}\tilde{\mathbf{E}}_{i,r}^{\dagger}\tilde{\mathbf{U}}_{i,k}^{(S)}\tilde{\mathbf{E}}_{i,r}+\mathbf{\Omega}_{i,r}^{(S)},\,\sum_{k\in\mathcal{N}_{U}}\tilde{\mathbf{E}}_{i,r}^{\dagger}\tilde{\mathbf{U}}_{i,k}^{(S)\prime}\tilde{\mathbf{E}}_{i,r}+\mathbf{\Omega}_{i,r}^{(S)\,\prime}\right)\nonumber \\
 & -\log_{2}\det\left(\mathbf{\Omega}_{i,r}^{(S)}\right),\\
\hat{\gamma}_{i,r}^{(S)}\left(\tilde{\mathbf{U}},\mathbf{\Sigma},\tilde{\mathbf{U}}^{\prime},\mathbf{\Sigma}^{\prime}\right)= & \frac{1}{\ln2}\varphi\left(\sum_{k\in\mathcal{N}_{U}}\bar{\mathbf{E}}_{\bar{i},r}^{\dagger}\tilde{\mathbf{U}}_{\bar{i},k}^{(S)}\bar{\mathbf{E}}_{\bar{i},r}+\mathbf{\Sigma}_{i,r}^{(S)},\,\sum_{k\in\mathcal{N}_{U}}\bar{\mathbf{E}}_{\bar{i},r}^{\dagger}\tilde{\mathbf{U}}_{\bar{i},k}^{(S)\prime}\bar{\mathbf{E}}_{\bar{i},r}+\mathbf{\Sigma}_{i,r}^{(S)\,\prime}\right)\nonumber \\
 & -\log_{2}\det\left(\mathbf{\Sigma}_{i,r}^{(S)}\right),\\
\hat{\beta}_{i,k,S}\left(\tilde{\mathbf{U}},\mathbf{\Omega},\tilde{\mathbf{U}}^{\prime},\mathbf{\Omega}^{\prime}\right)= & \frac{1}{\ln2}\varphi\left(\sum_{l\in\mathcal{N}_{U}}\bar{\mathbf{E}}_{i}^{\dagger}\tilde{\mathbf{U}}_{i,k}^{(S)}\bar{\mathbf{E}}_{i}+\mathbf{\Sigma}_{\bar{i}}^{(S)},\,\sum_{l\in\mathcal{N}_{U}}\bar{\mathbf{E}}_{i}^{\dagger}\tilde{\mathbf{U}}_{i,k}^{(S)\prime}\bar{\mathbf{E}}_{i}+\mathbf{\Sigma}_{\bar{i}}^{(S)\,\prime}\right)\nonumber \\
 & -\log_{2}\det\left(\sum_{l\in\mathcal{N}_{U}\setminus\{k\}}\bar{\mathbf{E}}_{i}^{\dagger}\tilde{\mathbf{U}}_{i,k}^{(S)}\bar{\mathbf{E}}_{i}+\mathbf{\Sigma}_{\bar{i}}^{(S)}\right).
\end{align}
with the notations $\varphi(\mathbf{A},\mathbf{B})=\ln\det(\mathbf{B})+\mathrm{tr}(\mathbf{B}^{-1}(\mathbf{A}-\mathbf{B}))$,
$\mathbf{G}_{i,k}^{j}=[\mathbf{H}_{i,k}^{j}\,\mathbf{H}_{i,k}^{\bar{j}}]$,
$\tilde{\mathbf{E}}_{i,r}=[\mathbf{E}_{i,r}^{\dagger}\,\mathbf{0}_{n_{R,\bar{i}}\times n_{R,i,r}}^{\dagger}]^{\dagger}$,
$\bar{\mathbf{E}}_{i,r}=[\mathbf{0}_{n_{R,i}\times n_{R,\bar{i},r}}^{\dagger}\,\mathbf{E}_{\bar{i},r}^{\dagger}]^{\dagger}$
and $\bar{\mathbf{E}}_{i}=[\mathbf{0}_{n_{R,i}\times n_{R,\bar{i}}}^{\dagger}\,\mathbf{I}_{n_{R,\bar{i}}}]^{\dagger}$.


\begin{thebibliography}{15}
\bibitem{Khan-et-al}
A. Khan, W. Kellerer, K. Kozu and M. Yabusaki, "Network sharing in the next mobile network: TCO reduction, management flexibility, and operational independence," \textit{IEEE Commn. Mag.}, vol. 49, no. 10, pp. 134-142, Oct. 2011.
\bibitem{Jorswieck-et-al}
E. A. Jorswieck, L. Badia, T. Fahldieck, E. Karipidis and J. Luo, "Spectrum sharing improves the network efficiency for cellular operators," \textit{IEEE Commun. Mag.}, vol. 52, no. 3, pp. 129-136, Mar. 2014.
\bibitem{Samdanis-et-al}
K. Samdanis, X. Costa-Perez and V. Sciancalepore, "From network sharing to multi-tenancy: The 5G network slicer broker," \textit{IEEE Commun. Mag.}, vol. 54, no. 7, pp. 32-39, Jul. 2016.
\bibitem{Boccardi-et-al}
F. Boccardi, H. Shorkri-Ghadikolaei, G. Fodor, E. Erkip, G. Fischione, M. Kountouris, P. Popovski and M. Zorzi, "Spectrum pooling in mmWave networks: Opportunities, challenges, and enablers," \textit{IEEE Commun. Mag.}, vol. 54, no. 11, pp. 33-39, Nov. 2016.
\bibitem{Aydin-et-al}
O. Aydin, E. A. Jorswieck, D. Aziz and A. Zappone, "Energy-spectral efficiency tradeoffs in 5G multi-operator networks with heterogeneous constraints," \textit{IEEE Trans. Wireless Comm.}, vol. 16, no. 9, pp. 5869-5881, Sep. 2017.
\bibitem{JPark-et-al}
J. Park, J. G. Andrews and R. W. Heath Jr., "Inter-operator base station coordination in spectrum-shared millimeter wave cellular networks," arXiv:1709.06239, Sep. 2017.
\bibitem{Foukas-et-al}
X. Foukas, G. Patounas, A. Elmokashfi and M. K. Marina, "Network slicing in 5G: Survey and challenges," \textit{IEEE Comm. Mag.}, vol. 55, no. 5, pp. 94-100, May 2017.
\bibitem{Simeone-et-al:JCN}
O. Simeone, A. Maeder, M. Peng, O. Sahin and W. Yu, "Cloud radio access network: Virtualizing wireless access for dense heterogeneous systems," \textit{Journ. Comm. Networks}, vol. 18, no. 2, pp. 135-149, Apr. 2016.
\bibitem{Quek-et-al}
T. Q. Quek, M. Peng, O. Simeone and W. Yu, \textit{Cloud Radio Access Networks: Principles, Technologies, and Applications}, Cambridge Univ. Press, Apr. 2017.
\bibitem{Simeone-et-al:ETT}
O. Simeone, O. Somekh, H. V. Poor and S. Shamai (Shitz), "Downlink multicell processing with limited-backhaul capacity," \textit{EURASIP J. Adv. Sig. Proc.}, 2009.
\bibitem{Park-et-al:TSP13}
S.-H. Park, O. Simeone, O. Sahin and S. Shamai (Shitz), "Joint precoding and multivariate backhaul compression for the downlink of cloud radio access networks," \textit{IEEE Trans. Sig. Processing}, vol. 61, no. 22, pp. 5646-5658, Nov. 2013.
\bibitem{Park-et-al:SPM}
S.-H. Park, O. Simeone, O. Sahin and S. Shamai (Shitz), "Fronthaul compression for cloud radio access networks: Signal processing advances inspired by network information theory," \textit{IEEE Sig. Proc. Mag.}, vol. 31, no. 6, pp. 69-79, Nov. 2014.
\bibitem{Tao-et-al}
M. Tao, E. Chen, H. Zhou and W. Yu, "Content-centric sparse multicast beamforming for cache-enabled cloud RAN," \textit{IEEE Trans. Wireless Comm.}, vol. 15, no. 9, pp. 6118-6131, Sep. 2016.
\bibitem{Lee-et-al:TSP16}
W. Lee, O. Simeone, J. Kang and S. Shamai (Shitz), "Multivariate fronthaul quantization for downlink C-RAN," \textit{IEEE Trans. Sig. Proc.}, vol. 64, no. 19, pp. 5025-5037, Oct. 2016.
\bibitem{Park-et-al:TWC}
S.-H. Park, O. Simeone and S. Shamai (Shitz), "Joint optimization of cloud and edge processing for fog radio access networks," \textit{IEEE Trans. Wireless Comm.}, vol. 15, no. 11, pp. 7621-7632, Nov. 2016.
\bibitem{Liu-Yu}
L. Liu and W. Yu, "Cross-layer design for downlink multihop cloud radio access networks with network coding," \textit{IEEE Trans. Sig. Proc.}, vol. 65, no. 7, pp. 1728-1740, Apr. 2017.
\bibitem{Park-et-al:TVT13}
S.-H. Park, O. Simeone, O. Sahin and S. Shamai (Shitz), "Robust and efficient distributed compression for cloud radio access networks," \textit{IEEE Trans. Veh. Tech.}, vol. 62, no. 2, pp. 692-703, Feb. 2013.
\bibitem{Zhou-et-al:TIT16}
Y. Zhou, Y. Xu, W. Yu and J. Chen, "On the optimal fronthaul compression and decoding strategies for uplink cloud radio access networks," \textit{IEEE Trans. Inf. Theory}, vol. 62, no. 2, pp. 7402-7418, Dec. 2016.
\bibitem{He-Yener}
X. He and A. Yener, "Cooperation with an untrusted realy: A secrecy perspective," \textit{IEEE Trans. Inf. Theory}, vol. 56, no. 8, pp. 3807-3827, Aug. 2010.
\bibitem{Bassily-et-al}
R. Bassily, E. Ekrem, X. He, E. Tekin, J. Xie, M. R. Bloch, S. Ulukus and A. Yener, "Cooperative security at the physical layer," \textit{IEEE Sig. Proc. Mag.}, vol. 30, no. 5, pp. 16-28, Sep. 2013.
\bibitem{Ostergaard-Zamir}
J. Ostergaard and R. Zamir, "Multiple-description coding by dithered delta-sigma quantization," \textit{IEEE Trans. Inf. Theory}, vol. 55, no. 10, pp. 4661-4675, Oct. 2009.
\bibitem{ElGamal-Kim}
A. E. Gamal and Y.-H. Kim, \textit{Network Information Theory}, Cambridge University Press, 2011.
\bibitem{Csiszar-Korner}
I. Csiszar and J. K. Korner, \textit{Information Theory: Coding Theorems for Discrete Memoryless Systems}, Academic Press, London, 1981.
\bibitem{CVX}
M. Grant and S. Boyd, "CVX: Matlab software for disciplined convex programming," ver 2.0 beta, Sep. 2013. [Online]. Available: http://cvxr.com/cvx.
\end{thebibliography}
\end{document}